\documentclass[preprint,prd,aps,showpacs,showkeys,nofootinbib]{revtex4}
\usepackage{graphicx}
\usepackage{makecell}
\usepackage{bm}
\usepackage{url}
\usepackage{float}
\usepackage{subfigure}
\begin{document}


\title{The rare decays $h^0\rightarrow Z\gamma,V Z$ in the NB-LSSM}

\author{Xing-Xing Dong$^{1,2,3,4}$\footnote{dongxx@hbu.edu.cn},
Cai Guo$^{1,2,3}$,
Wen Lu$^{1,2,3}$,
Shu-Min Zhao$^{1,2,3}$\footnote{zhaosm@hbu.edu.cn},
Tai-Fu Feng$^{1,2,3,5}$\footnote{fengtf@hbu.edu.cn}}
\affiliation{$^1$ College of Physics Science $\&$ Technology, Hebei University, Baoding, 071002, China\\
$^2$ Hebei Key Laboratory of High-precision Computation and Application of Quantum Field Theory, 071002, China\\
$^3$ Hebei Research Center of the Basic Discipline for Computational Physics, Baoding, 071002, China\\
$^4$ Departamento de F\'{i}sica and CFTP, Instituto Superior T\'{e}cnico, Universidade de Lisboa,
Av. Rovisco Pais 1, 1049-001 Lisboa, Portugal\\
$^5$ Department of Physics, Chongqing University, Chongqing, 401331, China}

\begin{abstract}
This study investigates the Higgs rare decays $h^0\rightarrow Z\gamma,V Z$ within the next to minimum B-L supersymmetric model(NB-LSSM), where $V$ denotes a vector meson $(\phi,J/\psi,\Upsilon(1S),\rho^0,\omega)$. Compared with the minimal supersymmetric Standard Model(MSSM), the NB-LSSM introduces three singlet Higgs superfields, which mix with the Higgs doublets, thereby modifying the lightest Higgs boson mass and its couplings. The loop-induced contributions resulting from the effective $h^0Z\gamma$ coupling can generate new physics(NP) contributions, significantly affecting the theoretical predictions for these rare decays through the new parameters such as $\kappa$, $\lambda$, $\lambda_2, \cdots$. The results of this work provide a useful reference for probing NP beyond the Standard Model(SM).
\end{abstract}

\keywords{Higgs boson, rare Higgs decay, supersymmetric model}
\pacs{14.80.Da, 13.66.Fg, 12.60.Jv}

\maketitle

\section{Introduction\label{sec1}}
In 2012, the CMS and ATLAS Collaborations discovered the Higgs boson, completing the last piece of the puzzle in the SM\cite{h0CMS,h0ATLAS}. Since then, extensive theoretical and experimental studies have been carried out on the couplings between the Higgs boson and fermions, as well as gauge bosons, leading to a deeper understanding of the Higgs mechanism and electroweak symmetry breaking. However, the SM contains only a single spin-zero Higgs particle, and its structure still faces several theoretical challenges, such as the naturalness problem and the strong CP problem\cite{Naturalness,CPproblem}. As a result, many NP frameworks beyond the SM, such as the Two-Higgs-Doublet Model(2HDM)\cite{2HDM1,2HDM2}, supersymmetric models(SUSY)\cite{Supersymmetry1,Supersymmetry2}, and composite Higgs models\cite{Composite Higgs Models}, predict the existence of additional Higgs bosons.

In these extended models, the coupling structure between the Higgs and other particles can be significantly modified\cite{hZgSM2,hZgloopMSSM}. For instance, the Higgs-fermion interactions may exhibit flavor-changing neutral currents(FCNCs), or introduce new sources of CP violation, thereby making the nature of Higgs interactions more complex. Such effects would manifest as deviations in Higgs couplings relative to the SM predictions. Therefore, any deviations observed in precision measurements of Higgs couplings could serve as indirect signals of NP, carrying important theoretical and experimental implications.

As one of the extensions of the SM, the next to minimum B-L supersymmetric model(NB-LSSM)\cite{NB-LSSM1,NB-LSSM2,mB-L1,B-LSSM2,B-LSSM3,B-LSSM4} can naturally solve the $\mu$ problem of the MSSM\cite{MSSM1,MSSM2,MSSM3,MSSM4} through the additional singlet Higgs state $S$. Moreover, in contrast to the Higgs sector of the MSSM, the NB-LSSM contains three Higgs singlets $\eta$, $\bar{\eta}$ and $S$. Their mixing with the Higgs doublets, together with new interactions involving additional charged scalars and fermions, can induce corrections to the effective Higgs couplings. These effects may lead to sizable deviations from the SM predictions. Therefore, it is of paramount importance to probe Higgs couplings in all conceivable ways.

In our previous work, the SM-like Higgs mass and decay modes $h^0\rightarrow\gamma\gamma,WW^*,ZZ,f\bar f$ with $f=b,\tau$ in the NB-LSSM have been studied\cite{LFV Higgs decay}. In this work, we focus on the rare Higgs decays $h^0\rightarrow Z\gamma,VZ$, where $V$ denotes a vector meson $(\phi,J/\psi,\Upsilon(1S),\rho^0,\omega)$. In the SM, the branching fraction of the decay $h^0\rightarrow Z\gamma$ is expected to be relatively small, $(1.5\pm0.1)\times10^{-3}$, for the SM-like Higgs boson\cite{hZgSM1,hZgSM2}. Since there is no tree-level $h^0Z\gamma$ coupling, this process is loop-induced and thus provides a sensitive probe of NP effects\cite{hZgoneloop1,hZgoneloop2}. For the processes $h^0\rightarrow VZ$, there are two types of decay topologies\cite{hZV1,hZV2,hZV3,hZV4,hZV5}. One is the direct contribution, in which the Higgs boson couples directly to the quarks forming the vector meson. The other is the indirect contribution, in which the Higgs couples to an off-shell vector boson that subsequently hadronizes into the vector meson through a local matrix element. The indirect contributions, arising from the effective $h^0Z\gamma$ coupling, are generally more sensitive to NP effects than the direct contributions. Therefore, it is necessary to study the indirect contributions mediated by the effective $h^0Z\gamma$ vertex. In addition, the QCD factorization approach is employed to analyze the processes $h^0\rightarrow VZ$\cite{QCDfactorization1,QCDfactorization2,QCDfactorization3,QCDfactorization4}.

Experimentally, the first evidence for the decay $h^0\rightarrow Z\gamma$ has been reported by the ATLAS and CMS Collaborations using data corresponding to an integrated luminosity of about $140\mathrm{fb}^{-1}$ at a center-of-mass energy of 13 TeV\cite{PDG2024,hZgexp}. The measured signal strength is $2.2\pm0.7$ times the SM prediction, with a statistical significance of 3.4 standard deviations. Nevertheless, a more recent ATLAS analysis\cite{hZgexp2025} reports a signal strength of $1.3^{+0.6}_{-0.5}$, corresponding to an observed significance of 2.5 standard deviations, which is closer to the SM expectation. So far, no evidence has been observed for Higgs decays into a $Z$ boson and a vector meson, although upper limits have been placed on several channels. The first experimental limits on the decay channels $h^0\rightarrow Z\rho^0,Z\phi$ were obtained by the CMS Collaboration using proton-proton collision data at $\sqrt{s} = 13\mathrm{TeV}$ with an integrated luminosity of $137\mathrm{fb}^{-1}$\cite{PDG2024,hZVexp1}. The corresponding upper limits on the branching fractions at the $95\%$ confidence level(CL) are in the ranges $1.04-1.31\%$ and $0.31-0.40\%$, respectively, which are about 740-940 and 730-950 times larger than the SM expectations. In addition, the CMS Collaboration has set an upper limit on the branching fraction of decay $h^0\rightarrow ZJ/\psi$ at the $95\%$ CL to be $1.9\times10^{-3}$, based on data at $\sqrt{s} = 13\mathrm{TeV}$ with an integrated luminosity of $137\mathrm{fb}^{-1}$\cite{PDG2024,hZVexp2}. With the improvement of experimental sensitivity, these rare Higgs decays may be observed at the furture High-Luminosity LHC(HL-LHC), providing increasing opportunities to search for NP.

This work is organized as follows. In Sec.II, we briefly introduce the NB-LSSM. In Sec.III, we derive the signal strengths of rare Higgs decays $h^0\rightarrow Z\gamma,VZ$. The numerical analyses are presented in Sec.IV, and our conclusions are given in Sec.V. The lengthy formulae are collected in the Appendix.
\section{The NB-LSSM}
The local gauge group of the NB-LSSM is $SU(3)_C\otimes{SU(2)_L}\otimes{U(1)_Y}\otimes{U(1)_{B-L}}$, where $B$ and $L$ denote the baryon and lepton numbers, respectively. Compared with the MSSM, the NB-LSSM adds three singlet Higgs superfields $\hat{\eta}$, $\hat{\bar{\eta}}$, and $\hat{S}$, as well as three generations of right-handed neutrino superfields $\hat{\nu}_i$. These additional fields can enlarge the available parameter space under constraints from LEP, Tevatron and the LHC, and thus help to alleviate the hierarchy problem of the MSSM\cite{B-L hierarchy1,B-L hierarchy2}. The NB-LSSM is implemented in the package SARAH\cite{SARAH1,SARAH2,SARAH3}, which is used to generate the mass matrices and interaction vertices. The newly introduced chiral superfields and quantum numbers of the NB-LSSM are listed in the TABLE \ref{NB-Lquantum numbers}.
\begin{table}[t]
\caption{ \label{NB-Lquantum numbers}  The chiral superfields and quantum numbers in the NB-LSSM.}
\footnotesize
\begin{tabular}{|c|c|c|c|c|c|c|}
\hline
Superfield & Spin 0 & Spin \(\frac{1}{2}\) & Generations &\(U(1)_Y\otimes\, \text{SU}(2)_L\otimes\, \text{SU}(3)_C\otimes\, U(1)_{B-L}\) \\
\hline
\(\hat{\nu}_i\) & \(\tilde{\nu}_i\) & \(\nu_i\) & 3 & \((0,{\bf 1},{\bf 1},\frac{1}{2}) \) \\
\(\hat{\eta}\) & \(\eta\) & \(\tilde{\eta}\) & 1 & \((0,{\bf 1},{\bf 1},-1) \) \\
\(\hat{\bar{\eta}}\) & \(\bar{\eta}\) & \(\tilde{\bar{\eta}}\) & 1 & \((0,{\bf 1},{\bf 1},1) \) \\
\(\hat{S}\) & \(S\) & \(\tilde{S}\) & 1 & \((0,{\bf 1},{\bf 1},0) \) \\
\hline
\end{tabular}
\end{table}

The superpotential of the NB-LSSM is given by
\begin{eqnarray}
&&W=Y_u^{ij}\hat U_i\hat Q_j\hat H_u-Y_d^{ij}\hat D_i\hat Q_j\hat H_d-Y_e^{ij}\hat E_i\hat L_j\hat H_d+Y_x^{ij}\hat{\nu}_i\hat{\eta}\hat{\nu}_j+Y_\nu^{ij}\hat{L}_i\hat{H}_u\hat{\nu}_j
\nonumber\\&&\hspace{0.6cm}+\lambda \hat S\hat H_u\hat H_d-\lambda_2 \hat S {\hat\eta}\hat{\bar{\eta}}+\frac{1}{3}\kappa \hat S\hat S\hat S,
\label{Wb}
\end{eqnarray}
where $i,j$ are generation indices, $Y_{u}^{ij},Y_{d}^{ij},Y_{e}^{ij},Y_{x}^{ij},Y_{\nu}^{ij},\lambda,\lambda_2$ and $\kappa$ denote the Yukawa and Higgs-sector couplings. After the Higgs doublets $H_d,H_u$ and the Higgs singlets $\eta,\bar{\eta},S$ acquire nonzero vacuum expectation values(VEVs) $v_d,v_u$ and $v_\eta,v_{\bar{\eta}},v_S$, respectively, the gauge group $SU(2)_L\otimes U(1)_Y\otimes U(1)_{B-L}$ is spontaneously broken down to $U(1)_{em}$. In this framework, tiny neutrino masses are generated at tree level through type-I seesaw mechanism. Meanwhile, the effective $\mu$ parameter, $\mu=\lambda v_S/\sqrt{2}$, is generated dynamically, thereby providing a natural solution to the $\mu$ problem. The Higgs fields are expanded around their VEVs as
\begin{eqnarray}
&&H_d=\left(
    \begin{array}{cc}
        H_d^- & \\
        \frac{1}{\sqrt{2}}\Big( \phi_d + v_d+ i\sigma_d\Big) & \\
    \end{array}
\right) , \quad  H_u=\left(
    \begin{array}{cc}
        \frac{1}{\sqrt{2}}\Big (\phi_u + v_u + i \sigma_u\Big) & \\
        H_u^+ & \\
    \end{array}
\right),\nonumber\\&&\eta = \frac{1}{\sqrt{2}} \Big( \phi_{\eta} + v_{\eta} + i \sigma_{\eta}\Big )  , \;
\bar{\eta} =\frac{1}{\sqrt{2}} \Big( \phi_{\bar{\eta}}+ v_{\bar{{\eta}}} + i \sigma_{\bar{{\eta}}}\Big),\;
S = \frac{1}{\sqrt{2}} \Big( \phi_S+ v_S + i \sigma_S\Big),
\end{eqnarray}
with $u^2=v_{\eta}^2+v_{\bar{\eta}}^2,\;v^2=v_d^2+v_u^2$ and $\tan\beta'=\frac{v_{\bar{\eta}}}{v_\eta}$, $\tan\beta=\frac{v_u}{v_d}$.

Additionally, the SUSY soft-breaking terms of the NB-LSSM can be written as
\begin{eqnarray} &&\mathcal{L}_{soft}=\mathcal{L}_{soft}^{MSSM}-\frac{T_\kappa}{3}S^3+T_{\lambda}SH_dH_u+T_{2}S\eta\bar\eta
-T_{\chi,ij}\eta\tilde{\nu}_{R,i}^{*}\tilde{\nu}_{R,j}^{*}-T_{\nu,ij}H_u\tilde{\nu}_{R,i}^{*}\tilde{\nu}_{L,j}
\nonumber\\&&\hspace{1cm}-m_{\eta}^2|\eta|^2\hspace{-0.05cm}-\hspace{-0.05cm}m_{\bar{\eta}}^2|\bar{\eta}|^2\hspace{-0.05cm}-\hspace{-0.05cm}m_S^2|S|^2\hspace{-0.05cm}-\hspace{-0.05cm}m_{\nu,ij}^2\tilde{\nu}_{R,i}^{*}\tilde{\nu}_{R,j}	\hspace{-0.05cm}-\hspace{-0.05cm}\frac{1}{2}(2M_{BB^\prime}\tilde{B}\tilde{B^\prime}\hspace{-0.05cm}+\hspace{-0.05cm}M_{BL}\tilde{B^\prime}^2\hspace{-0.05cm}+\hspace{-0.05cm}h.c.).
\end{eqnarray}
Here, $\mathcal{L}_{soft}^{MSSM}$ corresponds to the soft-breaking terms of the MSSM. The parameters $T_{\kappa}$, $T_{\lambda}$, $T_2$, $T_{\chi}$ and $T_{\nu}$ are the trilinear soft-breaking couplings with mass dimension. The fields $\tilde{B},\tilde{B^\prime}$ denote the gauginos of ${U(1)_Y}$ and ${U(1)_{B-L}}$, which mix with the $SU(2)_L$ gaugino $\tilde{W}$ and Higgsino $\tilde {H}_d,\tilde{H}_u, \tilde{\eta},\tilde{\bar{\eta}},\tilde{ S}$ to form Majorana neutralinos. The lightest neutralino can serve as a dark matter(DM) candidate, implying that the NB-LSSM offers a richer DM sector than the MSSM\cite{B-LDM1,B-LDM2,B-LDM3,B-LDM4}.

Compared with the MSSM and other SUSY models, the coexistence of the two Abelian groups ${U(1)_Y}$ and ${U(1)_{B-L}}$ in the NB-LSSM induces a new phenomenon known as gauge kinetic mixing. As a result, the hypercharge coupling $g_1$ is modified in the NB-LSSM, and the gauge sector is characterized by three couplings: $g_1$, the ${U(1)_{B-L}}$ gauge coupling $g_B$ and the mixing coupling $g_{YB}$, associated with mixing between the $Z$ boson and the new gauge boson $Z'$\cite{gBgYBB-L}. This gauge kinetic mixing also leads to Higgs mixing and gaugino-higgsino mixing at tree level, thereby affecting the mass matrices of the Higgs bosons, neutralinos, squarks and sleptons. Consequently, these effects can significantly influence the Higgs mass spectrum and the predictions for Higgs decays in the NB-LSSM.

Additionally, the tree-level Higgs mass-squared matrix in the basis $(\phi_d,\phi_u,\phi_{\eta},\phi_{\bar{\eta}},\phi_S) $ can be deduced as
\begin{eqnarray}
&&m_{h} ^2 = \left(
	\begin{array}{cccccccc}
		m_{\phi_d\phi_d}  &m_{\phi_u\phi_d} &m_{\phi_{\eta}\phi_d} &m_{\phi_{\bar{\eta}}\phi_d} &m_{\phi_S\phi_d}\\
		m_{\phi_d\phi_u}  &m_{\phi_u\phi_u} &m_{\phi_{\eta}\phi_u} &m_{\phi_{\bar{\eta}}\phi_u} &m_{\phi_S\phi_u}\\
		m_{\phi_d\phi_{\eta}}  &m_{\phi_u\phi_{\eta}} &m_{\phi_{\eta}\phi_{\eta}} &m_{\phi_{\bar{\eta}}\phi_{\eta}} &m_{\phi_S\phi_{\eta}}\\
		m_{\phi_d\phi_{\bar{\eta}}}  &m_{\phi_u\phi_{\bar{\eta}}} &m_{\phi_{\eta}\phi_{\bar{\eta}}} &m_{\phi_{\bar{\eta}}\phi_{\bar{\eta}}} &m_{\phi_S\phi_{\bar{\eta}}}\\
		m_{\phi_d\phi_S}  &m_{\phi_u\phi_S} &m_{\phi_{\eta}\phi_S} &m_{\phi_{\bar{\eta}}\phi_S} &m_{\phi_S\phi_S}\\
	\end{array}
	\right),\nonumber\\&&
m_{\phi_d\phi_d} = \frac{1}{4} G^2 v_d^2 - L_{\phi_d\phi_u} \tan\beta,\;\; m_{\phi_u\phi_u} =\frac{1}{4} G^2 v_u^2 - L_{\phi_d\phi_u} \cot\beta, \nonumber\\&&
m_{\phi_{\eta}\phi_{\eta}} = g_B^2 v_{\eta}^2 - L_{\phi_{\eta}\phi_{\bar{\eta}}} \tan\beta' ,\;\;m_{\phi_{\bar{\eta}}\phi_{\bar{\eta}}} = g_B^2 v_{\bar{\eta}}^2 - L_{\phi_{\eta}\phi_{\bar{\eta}}} \cot\beta', \nonumber\\&&
m_{\phi_d\phi_u} =-\frac{1}{4} (G^2-4|\lambda|^2) v_d v_u + L_{\phi_d\phi_u},\;\;   m_{\phi_{\eta}\phi_{\bar{\eta}}} = -(g_B^2-4|\lambda_2|^2) v_{\eta} v_{\bar{\eta}}+L_{\phi_{\eta}\phi_{\bar{\eta}}}, \nonumber\\&&
m_{\phi_d\phi_{\eta}} = \frac{g_1 + g_{YB} g_B}{2} v_d v_{\eta} + \frac{\Re(\lambda^* \lambda_2)}{2} v_u  v_{\bar{\eta}}  ,\;\;
m_{\phi_u\phi_{\eta}} = -\frac{g_1 +g_{YB} g_B}{2} v_u v_{\eta} + \frac{\Re(\lambda^* \lambda_2)}{2} v_d v_{\bar{\eta}},  \nonumber\\&&
m_{\phi_d\phi_{\bar{\eta}}} = -\frac{g_{YB} g_B}{2} v_d v_{\bar{\eta}} + \frac{\Re(\lambda^* \lambda_2)}{2} v_u v_{\eta}  ,\;\;
m_{\phi_u\phi_{\bar{\eta}}} = \frac{g_{YB} g_B}{2}v_u v_{\bar{\eta}}+ \frac{\Re(\lambda^* \lambda_2)}{2} v_d v_{\eta},  \nonumber\\&&
m_{\phi_d\phi_S} = \frac{1}{2} \left(- v_u \left( \lambda v_S \kappa^* + \sqrt{2} \Re(T_\lambda) \right) + v_S \left( 2 \lambda v_d - \kappa v_u \right) \lambda^* \right), \nonumber\\&&
m_{\phi_u\phi_S} = \frac{1}{2} \left( -v_d \left( \lambda v_S \kappa^*+ \sqrt{2}  \Re(T_\lambda) \right) - v_S \left( -2 \lambda v_u + \kappa v_d \right) \lambda^* \right),\nonumber\\&&
m_{\phi_{\eta}\phi_S} = \frac{1}{2} \left( -v_{\bar{\eta}} \left( \lambda_2 v_S \kappa^*+ \sqrt{2}  \Re(T_2) \right) + v_S \left( 2 \lambda_2 v_{\eta}-\kappa v_{\bar{\eta}} \right) \lambda_2^* \right),\nonumber\\&&
m_{\phi_{\bar{\eta}}\phi_S} = \frac{1}{2} \left( -v_{\eta} \left( \lambda_2 v_S \kappa^*+ \sqrt{2}  \Re(T_2) \right) - v_S \left( -2 \lambda_2 v_{\bar{\eta}}+\kappa v_{\eta} \right) \lambda_2^* \right),\nonumber\\&&
m_{\phi_S\phi_S} = 2 \kappa^* \kappa v_S^2 + \frac{\sqrt{2}}{2} \Re(T_\kappa) v_S + \frac{\sqrt{2}}{2} \Re(T_\lambda) \frac{v_u v_d}{v_S} + \frac{\sqrt{2}}{2} \Re(T_2) \frac{v_{\eta}v_{\bar{\eta}}} {v_S}.\label{higgs}
\end{eqnarray}
Here, $G^2=g_{1}^2+g_{2}^2+g_{YB}^2$, $L_{\phi_d\phi_u}=\frac{1}{4} [ -2\sqrt{2} v_S \Re(T_\lambda)+ (-\kappa v_S^2 + \lambda_2 v_{\eta} v_{\bar{\eta}} )\lambda^* + \lambda(v_{\eta} v_{\bar{\eta}} \lambda_2^* - v_S^2 \kappa^*) ]$ and $L_{\phi_{\eta}\phi_{\bar{\eta}}}=\frac{1}{4} [ -2\sqrt{2} v_S \Re(T_2) + (-\kappa v_S^2 + \lambda v_d v_u)\lambda_2^* + \lambda_2(v_d v_u \lambda^* - v_S^2 \kappa^*) ]$. The mass-squared matrix $m_{h} ^2$ is diagonalized by the rotation matrix $Z^H$ as $Z^H m_{h}^2 Z^{H \dagger} = (m_{h_i}^2)^{diag}$. Including the leading-log radiative corrections from the top/stop sector, the mass of the SM-like Higgs boson is approximated by $m_{h^0}=\sqrt{m_{h_1}^2+\Delta m_h^2}$, where $m_{h_1}$ denotes the lightest tree-level mass eigenstate. The leading-log radiative corrections $\Delta m_h^2$ is given by \cite{leadinglog1,leadinglog2,leadinglog3}
\begin{eqnarray}
&&\Delta m_h^2=\frac{3m_t^4}{4 \pi ^2 v^2}\Big[\Big(\tilde{t}+\frac{1}{2}\tilde{X}_t\Big)+\frac{1}{16 \pi ^2}\Big(\frac{3m_t^2}{2v^2}-32\pi\alpha_3\Big)(\tilde{t}^2+\tilde{X}_t\tilde{t})\Big],\nonumber \\&&\tilde{t}=\log\frac{M_S^2}{m_t^2},~~~~\tilde{X}_t=\frac{2\tilde{A}_t^2}{M_S^2}(1-\frac{\tilde{A}_t^2}{12M_S^2}).
\end{eqnarray}
Here, $\alpha_3$ is the strong coupling constant, and $M_S=\sqrt{m_{\tilde{t}_1}m_{\tilde{t}_2}}$ is the geometric mean of the stop masses $m_{\tilde{t}_{1,2}}$. The stop mixing parameter is $\tilde{A}_t=A_t-\frac{\lambda v_S }{\sqrt{2}}\cot \beta $ where $A_t=T_{u,33}$ denotes the trilinear Higgs-stops coupling.

\section{The rare Higgs decays $h^0\rightarrow Z\gamma,V Z$}
In this section, we present the analytical formulae of the rare Higgs decays $h^0\rightarrow Z\gamma,V Z$ with $V\in(\phi,J/\psi,\Upsilon(1S),\rho^0,\omega)$.
\begin{figure}[t]
\centering
\includegraphics[width=0.5\textwidth]{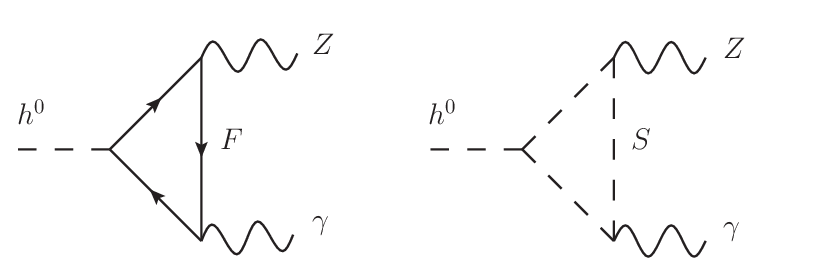}
\caption[]{The one-loop diagrams for decay $h^{0}\rightarrow Z\gamma$ in the NB-LSSM. Here $F$ and $S$ denote fermions and scalars, respectively.}
\label{fig1}
\end{figure}

The effective Lagrangian for process $h^{0}\rightarrow Z\gamma$ can be written as:
\begin{eqnarray}
&&\mathcal{L}_{\mathrm{eff}}=\frac{\alpha}{4 \pi v}\left(\frac{2 C_{\gamma Z}}{s_{w} c_{w}} h F_{\mu \nu} Z^{\mu \nu}-\frac{2 \tilde{C}_{\gamma Z}}{s_{w} c_{w}} h F_{\mu \nu} \tilde{Z}^{\mu \nu}\right),\label{hzr}
\end{eqnarray}
corresponding, $s_{w}=\sin \theta_{w}$ and $c_{w}=\cos \theta_{w}$ with $\theta_{w}$ being the Weinberg angle. Similar to the $h^0\gamma\gamma$ vertex, the $h^0Z\gamma$ interaction arises only at loop level. In the SM, the loop contributions are generated by the fermions and the W boson\cite{hZgloopSM}. In the NB-LSSM, additional particles such as charginos and sfermions can contribute, potentially inducing sizable NP effects in the effective $h^0Z\gamma$ coupling\cite{hZgloopMSSM}. The relevant one-loop diagrams contributing to the decay $h^0\rightarrow Z\gamma$ in the NB-LSSM are shown in Fig.\ref{fig1}. Using the effective Lagrangian in Eq.(\ref{hzr}), the decay width of the loop-induced process $h^0\rightarrow Z\gamma$ can be deduced as:
\begin{eqnarray}
\Gamma\left(h^{0} \rightarrow Z \gamma\right)=\frac{\alpha^{2} m_{h^{0}}^{3}}{32 \pi^{3} v^{2} s_{w}^{2} c_{w}^{2}}\left(1-\frac{m_{Z}^{2}}{m_{h^{0}}^{2}}\right)^{3}\left(\left|C_{\gamma Z}\right|^{2}+\left|\tilde{C}_{\gamma Z}\right|^{2}\right).
\end{eqnarray}
The explicit expressions of $ C_{\gamma Z} $ and $ \tilde{C}_{\gamma Z} $ can be written as\cite{hZgloopSM,hZgloopMSSM,hZgloopNMSSM,hZgloopChen}:
\begin{eqnarray}
&&C_{\gamma Z}^{\mathrm{SM}}= \sum_{q}  2N_{c} Q_{q} v_{q} A_{1/2}\left(\tau_{q}, \lambda_{q}\right)  +\sum_{l} 2Q_{l} v_{l} A_{1/2}\left(\tau_{l}, \lambda_{l}\right)+\frac{c_w}{2} A_{1}\left(\tau_{W},\lambda_{W}\right),\nonumber\\ &&C_{\gamma Z}^{\mathrm{NP}}= \sum_{q}  2N_{c} Q_{q}  g_{h^0qq}g_{Zqq} A_{1/2}\left(\tau_{q}, \lambda_{q}\right)  +\sum_{l} 2Q_{l} g_{h^0ll}g_{Zll} A_{1/2}\left(\tau_{l}, \lambda_{l}\right)\nonumber \\
&&\hspace{-0.2cm}+\sum_{\chi^{\pm}_i} Q_{\chi^{\pm}_i} \frac{m_W}{m_{\chi^{\pm}_i}}g_{h^0\chi^{\pm}_i\chi^{\pm}_i}g_{Z\chi^{\pm}_i\chi^{\pm}_i} A_{1/2}\left(\tau_{\chi^{\pm}_i}, \lambda_{\chi^{\pm}_i}\right)+ \frac{c_w}{2}g_{h^0WW}g_{ZWW}A_{1}\left(\tau_{W}, \lambda_{W}\right)
\nonumber \\
&&\hspace{-0.2cm}+\hspace{-0.1cm}\sum_{H^{\pm}_i}  \frac{Q_{H^{\pm}_i}m_W^2}{4c_w^2m_{H^{\pm}_i}^2}g_{h^0H^{\pm}_iH^{\pm}_i}g_{ZH^{\pm}_iH^{\pm}_i} A_{0}\left(\tau_{H^{\pm}_i}, \lambda_{H^{\pm}_i}\right)
\hspace{-0.1cm}-\hspace{-0.1cm}\sum_{\tilde{f}_i}  \frac{N_{c}Q_{\tilde{f}_i}m_W^2}{2m_{\tilde{f}_i}^2}g_{h^0\tilde{f}_i\tilde{f}_i}g_{Z\tilde{f}_i\tilde{f}_i} A_{0}\left(\tau_{\tilde{f}_i}, \lambda_{\tilde{f}_i}\right),
\end{eqnarray}
where $v_{q,l}=T_3^{q,l}/2-Q_{q,l}s_w^2$, $ \tau_{i}=\frac{4 m_{i}^{2}}{m_{h^0}^{2}} $, $ \lambda_{i}=\frac{4 m_{i}^{2}}{m_Z^{2}} $. The $C_{\gamma Z}^{\mathrm{SM}} $ corresponds to the SM contributions, while $C_{\gamma Z}^{\mathrm{NP}} $ corresponds to the new contributions from the NB-LSSM. In general, the contributions from non-diagonal couplings $g_{h^0\chi^{\pm}_i\chi^{\pm}_j}$, $g_{Z\chi^{\pm}_i\chi^{\pm}_j}$, $g_{h^0\tilde{f}_i\tilde{f}_j}$ and $g_{Z\tilde{f}_i\tilde{f}_j}$ are numerically small and can be neglected for most purposes\cite{hZgloopMSSM}. Therefore, we only consider the diagonal sfermion and chargino contributions in the loop. Moreover, in the SM one has $\tilde{C}_{\gamma Z}^{S M} \sim 0 $\cite{hZV1}, and we find $\tilde{C}_{\gamma Z}^{NP}$ to be also negligibly small in the NB-LSSM, hence the CP-odd contributions can be safely ignored. Besides, the loop functions $A_{1/2}$, $A_{1} $ and $A_{0}$ can be found in Appendix A, while the relevant couplings $g_{h^0qq}$, $g_{Zqq},\cdots$ are given in Appendix B.

\begin{figure}[t]
\centering
\includegraphics[width=0.8\textwidth]{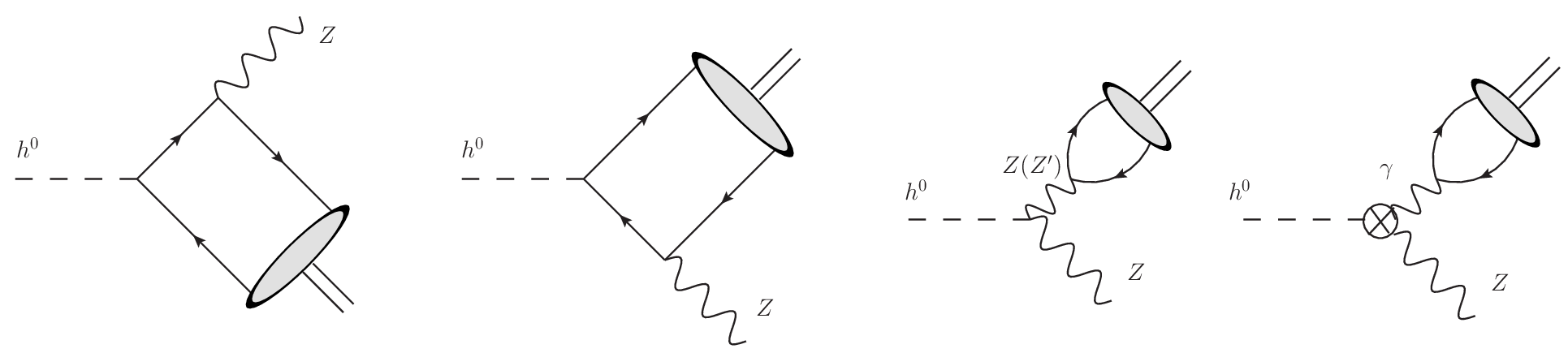}
\caption[]{The major diagrams for decay $h^{0}\rightarrow ZV$ in the NB-LSSM.}
\label{fig2}
\end{figure}
Next, we consider the Higgs boson weak hadronic decay $h^0 \rightarrow V Z$ with vector meson ${V}\in(\phi,J/\psi,\Upsilon(1S),\rho^0,\omega)$, as illustrated in Fig.\ref{fig2}. The first two diagrams represent the direct contributions, where the Higgs couples directly to the quark current forming the meson. The last two diagrams correspond to the indirect contributions, in which the meson is produced from an off-shell vector boson. In particular, the contribution induced by the loop-level $h^0 Z\gamma$ vertex in Fig.2(d) is especially sensitive to NP effects. The amplitudes of the decays $ h^{0} \rightarrow V Z $ can be parameterized as
\begin{eqnarray}
i \mathcal{A}(h^{0} \rightarrow V Z)=\frac{-2 g m_{V}}{c_{w} v}\Big[\varepsilon_{V}^{\| *} \cdot \varepsilon_{Z}^{\| *} F_{\|}^{V Z}+\varepsilon_{V}^{\perp *} \cdot \varepsilon_{Z}^{\perp *} F_{\perp}^{V Z} +\frac{\epsilon_{\mu \nu \alpha \beta} k_{V}^{\mu} k_{Z}^{\nu} \varepsilon_{V}^{* \alpha} \varepsilon_{Z}^{* \beta}}{(\left(k_{V} \cdot k_{Z}\right)^{2}-k_{V}^{2} k_{Z}^{2})^{1 / 2}} \tilde{F}_{\perp}^{V Z}\Big]
\end{eqnarray}
with $k_{Z}$ denoting the four-momentum of the $Z$ boson. The polarization vectors $\varepsilon_{V}^{\| \mu}$ and $\varepsilon_{V}^{\perp \mu}$ denote the longitudinal and transverse components of the vector meson polarization, respectively: $\varepsilon_{V}^{\| \mu} =\frac{1}{m_{V}} \frac{k_{V} \cdot k_{Z}}{[(k_{V} \cdot k_{Z})^{2}-k_{V}^{2} k_{Z}^{2}]^{1 / 2}}(k_{V}^{\mu}-\frac{k_{V}^{2}}{k_{V} \cdot k_{Z}} k_{Z}^{\mu})$, $\varepsilon_{V}^{\perp \mu}  =\varepsilon_{V}^{\mu}-\varepsilon_{V}^{\| \mu}$. The polarization vectors of the $ Z $ boson can be obtained by replacing $ m_{V} \rightarrow m_{Z}$ and interchanging $k_{V} \leftrightarrow k_{Z} $. $F_{\|}^{V Z}$ represents the CP-even longitudinal form factor, while $F_{\perp}^{V Z}$ and $\tilde{F}_{\perp}^{V Z}$ represent the CP-even and CP-odd transverse form factors, respectively. Both direct and indirect contributions enter these form factors. We first discuss the indirect contributions, which contain the loop-induced NP effects.
\begin{eqnarray}
&&F_{\| \text {indirect }}^{V Z}= \frac{g_{h^0ZZ}}{1-r_{V} / r_{Z}} \sum_{q} f_{V}^{q} g_{Zqq}+C_{\gamma Z} \frac{\alpha\left(m_{V}\right)}{4 \pi} \frac{4 r_{Z}}{1-r_{Z}-r_{V}} \sum_{q} f_{V}^{q} Q_{q},\nonumber \\
&&F_{\bot\text {indirect }}^{V Z}= \frac{g_{h^0ZZ}}{1-r_{V} / r_{Z}} \sum_{q} f_{V}^{q}  g_{Zqq} +C_{\gamma Z} \frac{\alpha\left(m_{V}\right)}{4 \pi} \frac{1-r_{Z}-r_{V}}{r_{V}} \sum_{q} f_{V}^{q} Q_{q}, \nonumber \\
&&\tilde{F}_{\bot\text {indirect }}^{V Z}= \tilde{C}_{\gamma Z} \frac{\alpha\left(m_{V}\right)}{4 \pi} \frac{\lambda^{1 / 2}\left(1, r_{Z}, r_{V}\right)}{r_{V}} \sum_{q} f_{V}^{q} Q_{q},
\end{eqnarray}
where the indirect contributions $\tilde{F}_{\bot\text {indirect }}^{V Z}$ can be ignored in sake of $\tilde{C}_{\gamma Z}\sim0$. Here, $r_{Z,V}=\frac{m_{Z,V}^{2}}{m_{h^{0}}^{2}}$, the flavor-specific decay constants $f_{V}^{q}$ are defined through the local matrix elements $\langle V(k, \varepsilon)| \bar{q} \gamma^{\mu} q|0\rangle=-i f_{V}^{q} m_{V} \varepsilon^{* \mu},\; q=u, d, s,\cdots$ together with the relations $\sum_{q} Q_{q} f_{V}^{q}=Q_{V} f_{V},\;\sum_{q} f_{V}^{q} g_{Zqq}=f_{V} g_{ZVV}$\cite{decay constant}. The concrete values of $f_{V}$, $Q_{V}$, $m_{V}$, $g_{ZVV}$ and the transverse decay-constant ratio $f_{V}^\top/f_{V}$ are listed in the TABLE \ref{meson}.
\begin{table}[t]
\caption{The values of $f_{V}$, $Q_{V}$, $m_{V}$, $g_{ZVV}$ and $f_{V}^\top$.}
\footnotesize
\begin{tabular}{|c|c|c|c|c|c|}
\hline
Meson $V$  &$m_{V}$/GeV&$Q_{V}$    & $g_{ZVV}$ &$f_{V}$& $f_{V}^\top/f_{V}$ \\
\hline
$\phi$        &1.019  &-1/3       & $g_{Zdd}$ &0.223  & 0.76 \\
\hline
$J/\psi$      &3.097  &2/3        & $g_{Zuu}$ &0.403  & 0.91 \\
\hline
$\Upsilon(1S)$&9.46   &-1/3       & $g_{Zdd}$ &0.648  & 1.09\\
\hline
$\rho^0$      &0.775  &$\frac{1}{\sqrt{2}}$ &$\frac{g_{{Zuu}}-g_{{Zdd}}}{\sqrt{2}}$ &0.216 &0.72\\
\hline
$\omega$      &0.782  &$\frac{1}{3\sqrt{2}}$&$\frac{g_{{Zuu}}+g_{{Zdd}}}{\sqrt{2}}$&0.194 &0.71\\
\hline
\end{tabular}\label{meson}
\end{table}

The direct contributions can be calculated in the power series of $\left(m_{q} / m_{h^0}\right)^{2} $ or $ \left(\Lambda_{Q C D} / m_{h^0}\right)^{2} $ comparing the indirect contributions, where $m_{q}$ corresponds to the constituent quark mass of a given meson and $\Lambda_{Q C D}$ is the hadronic scale. After using asymptotic function $\phi_{V}^{\perp}(x)=6 x(1-x)$, the concrete expression of direct contributions can be obtained
\begin{eqnarray}
&&F_{\perp \text { direct }}^{V Z}=\sum_{q} f_{V}^{q \perp} g_{h^0qq} g_{Zqq} \frac{3 m_{q}}{2 m_{V}} \frac{1-r_{Z}^{2}+2 r_{Z} \ln r_{Z}}{\left(1-r_{Z}\right)^{2}}, \nonumber \\
&&\tilde{F}_{\perp \text { direct }}^{V Z}=\sum_{q} f_{V}^{q \perp} \tilde{g}_{h^0qq} g_{Zqq} \frac{3 m_{q}}{2 m_{V}} \frac{1-r_{Z}^{2}+2 r_{Z} \ln r_{Z}}{\left(1-r_{Z}\right)^{2}}.
\end{eqnarray}
Numerically, we find that the direct contributions are strongly suppressed. Therefore, the indirect contributions-especially those involving the effective $h^0 Z\gamma$ vertex-play the dominant role in our analysis. The decay width of $ h^{0} \rightarrow V Z $ can then be written as
\begin{eqnarray}
&&\Gamma  \left(h^{0} \rightarrow V Z\right)=\frac{m_{h^{0}}^{3}}{4 \pi v^{4}} \lambda^{1 / 2}\left(1, r_{Z}, r_{V}\right)\left(1-r_{Z}-r_{V}\right)^{2} \nonumber \\
&& \hspace{2.9cm}\times\left[\left|F_{\|}^{V Z}\right|^{2}+\frac{8 r_{V} r_{Z}}{\left(1-r_{Z}-r_{V}\right)^{2}}\left(\left|F_{\perp}^{V Z}\right|^{2}+\left|\tilde{F}_{\perp}^{V Z}\right|^{2}\right)\right]
\end{eqnarray}
with $\lambda(x, y, z)=(x-y-z)^{2}-4 y z$.

The signal strengths for the channels $h^0\rightarrow Z\gamma,V Z$ are defined as \cite{signal strength ratios}
\begin{eqnarray}
&&\mu_{Z\gamma,VZ}^{ggF}=\frac{\sigma_{NP}(ggF)}{\sigma_{SM}(ggF)}
\frac{Br_{NP}(h^0\rightarrow Z\gamma,VZ)}{Br_{SM}(h^0\rightarrow Z\gamma,VZ)},(V\in(\phi,J/\psi,\Upsilon(1S),\rho^0,\omega)),
\end{eqnarray}
where $ggF$ stands for gluon-gluon fusion. The Higgs production cross sections can be approximated as
$\frac{\sigma_{NP}(ggF)}{\sigma_{SM}(ggF)}\approx
\frac{\Gamma_{NP}(h^0\rightarrow gg)}{\Gamma_{SM}(h^0\rightarrow gg)}$. Meanwhile, $Br_{NP}(\cdot\cdot\cdot)=\frac{\Gamma_{NP}(\cdot\cdot\cdot)}{\Gamma_{NP}^{h^0}}$ with $\Gamma_{NP}^{h^0}=\sum\Gamma_{NP}(h^0\rightarrow gg, \gamma\gamma, WW^*,ZZ,f\bar f,Z\gamma, VZ)$ denoting the NP total decay width of the SM-like Higgs boson. Analogously, $Br_{SM}(\cdot\cdot\cdot)=\frac{\Gamma_{SM}(\cdot\cdot\cdot)}{\Gamma_{SM}^{h^0}}$ with $\Gamma_{SM}^{h^0}$ denoting the SM total width. Therefore, the signal strengths of $h^0\rightarrow Z\gamma,V Z$ can be rewritten as
\begin{eqnarray}
&&\mu_{Z\gamma,VZ}^{ggF}\approx\frac{\Gamma_{SM}^{h^0}}{\Gamma_{NP}^{h^0}}\frac{\Gamma_{NP}(h^0\rightarrow gg)}{\Gamma_{SM}(h^0\rightarrow gg)}
\frac{\Gamma_{NP}(h^0\rightarrow Z\gamma,VZ)}{\Gamma_{SM}(h^0\rightarrow Z\gamma,VZ)}.
\end{eqnarray}

\section{The numerical analyses}
At present, the latest experimental data of SM-like Higgs boson mass and the corresponding signal strengths are $m_h^{0}=125.20\pm0.11$ GeV, $\mu_{\gamma\gamma}^{exp}=1.10\pm0.06$, $\mu_{WW^*}^{exp}=1.00\pm0.08$, $\mu_{ZZ}^{exp}=1.02\pm0.08$, $\mu_{b\bar{b}}^{exp}=0.99\pm0.12$, $\mu_{\tau\bar{\tau}}^{exp}=0.91\pm0.09$ and $\mu_{Z\gamma}^{exp}=1.3^{+0.6}_{-0.5}$\cite{PDG2024,hZgexp,hZgexp2025,h02gamma,h02gamma2W2Z2b2tau1,h02gamma2W2Z2b2tau2,h02gamma2W2b2tau,h02Z,h02b1,h02b2,h02tau}, which give the stringent constraints on parameter spaces of various SM extensions. Additionally, we apply the following experimental constraints:
(1) The updated searches require the $Z'$ boson mass to satisfy $M'_Z\geq 5.15~ {\rm TeV}$ at $95\%$ CL\cite{Zpupper}. An upper bound of the ratio between the $Z'$ boson mass and its gauge coupling is given out by Refs.\cite{Zpupper1,Zpupper2} $\frac{M'_Z}{g_B}\geq 6~{\rm TeV}$ at $99\%$ CL, which restricts the gauge coupling to $0 < g_B \leq0.85$. (2) The measured branching ratio of $\bar{B}\rightarrow X_s\gamma$ disfavors the large $\tan\beta$\cite{BSgamma1,BSgamma2}. (3) The LHC experimental data constrain $\tan\beta'< 1.5$\cite{tanB}. (4) The lower bounds on the masses of extended SM particles are considered: the slepton mass is greater than 0.7 TeV and the squark mass is greater than 2 TeV\cite{PDG2024}.

\begin{figure}[t]
\centering
\hspace{0cm}\includegraphics[width=0.4\textwidth]{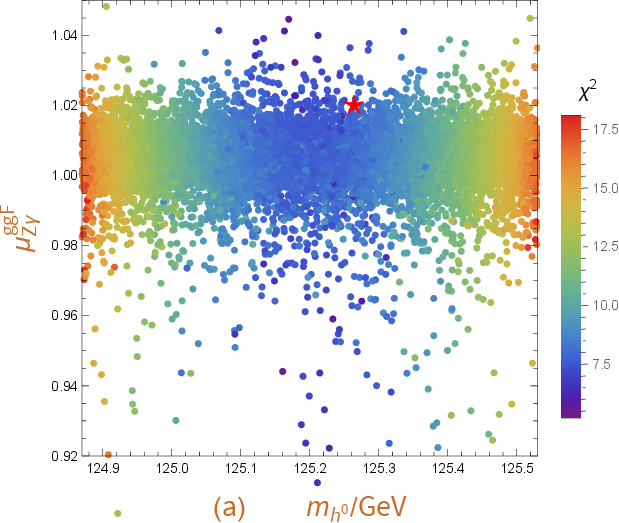}\\
\hspace{0cm}\includegraphics[width=0.4\textwidth]{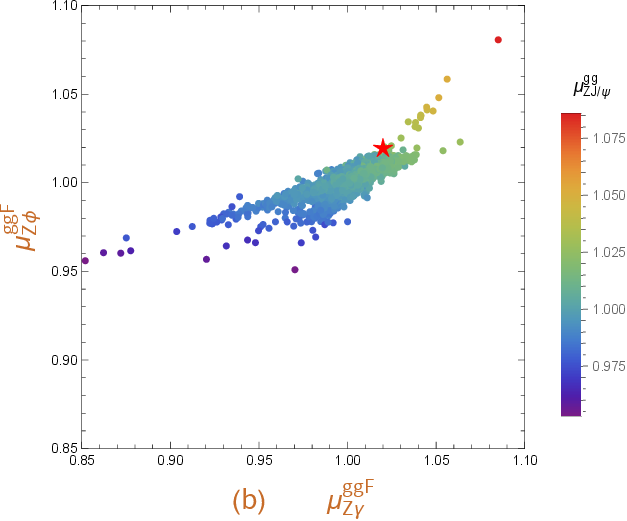}
\hspace{0.2cm}\includegraphics[width=0.4\textwidth]{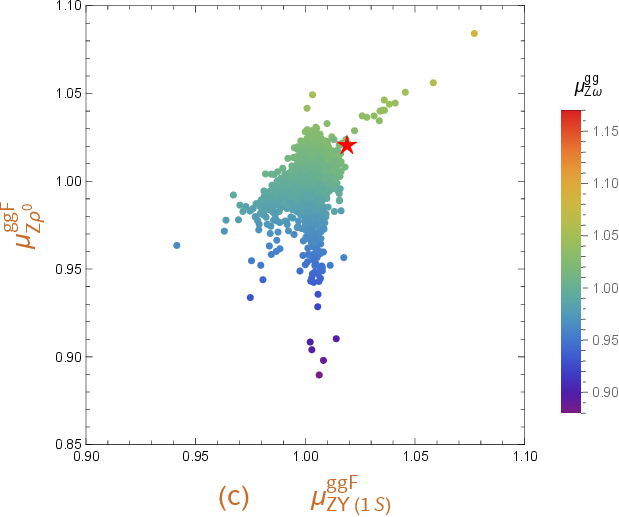}\\
\hspace{0cm}\includegraphics[width=0.38\textwidth]{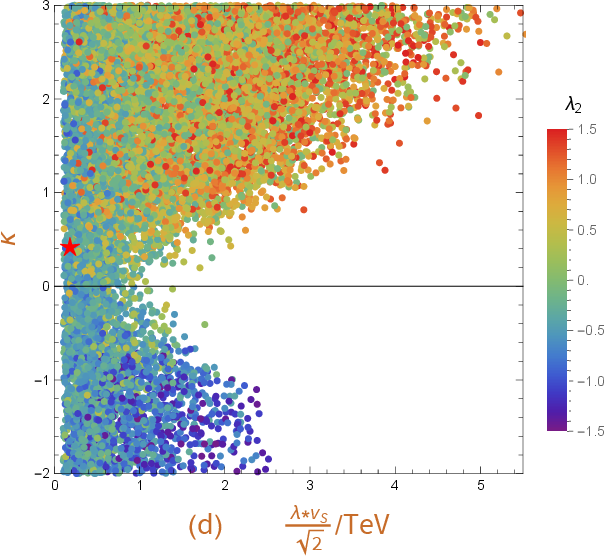}
\hspace{0.2cm}\includegraphics[width=0.4\textwidth]{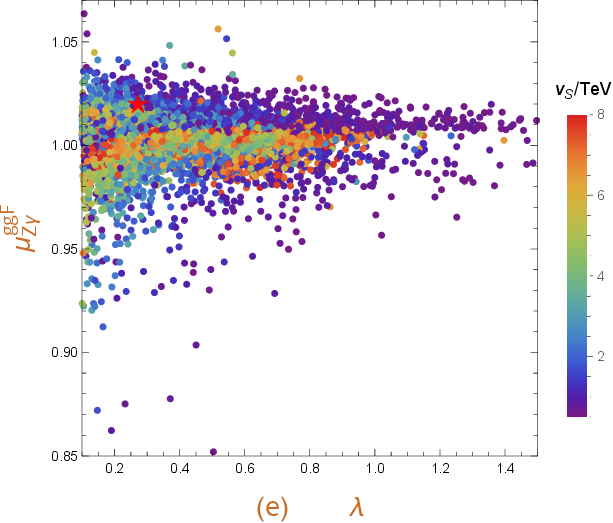}
\caption[]{The scatter points for parameters $\lambda,\kappa,\lambda_2,v_S$ and signal strength $\mu_{Z\gamma,ZV}^{ggF}$.}
\label{fig3}
\end{figure}

In order to analyze the signal strengths of rare decays $h^0\rightarrow Z\gamma,VZ$, we perform a $\chi^2$ fit to the Higgs mass and signal-strength data, in which the theoretical values obtained from our model $\mu_{\xi}^{theo}$ are confronted with the experimental measurements $\mu_{\xi}^{exp}$, $\delta_\xi$ represents the total uncertainty including both statistical and systematic.
\begin{eqnarray}
&&\chi^2=\sum_\xi(\frac{\mu_{\xi}^{theo}-\mu_{\xi}^{exp}}{\delta_\xi})^2
=(\frac{m_{h^0}^{theo}-m_{h^0}^{exp}}{\delta_{m_{h}^0}})^2
+(\frac{\mu_{\gamma\gamma}^{theo}-\mu_{\gamma\gamma}^{exp}}{\delta_{\mu_{\gamma\gamma}}})^2
+(\frac{\mu_{WW^*}^{theo}-\mu_{WW^*}^{exp}}{\delta_{\mu_{WW^*}}})^2
\nonumber\\
 &&\hspace{0.5cm}+(\frac{\mu_{ZZ}^{theo}-\mu_{ZZ}^{exp}}{\delta_{\mu_{ZZ}}})^2
+(\frac{\mu_{b\bar{b}}^{theo}-\mu_{b\bar{b}}^{exp}}{\delta_{\mu_{b\bar{b}}}})^2
+(\frac{\mu_{\tau\bar{\tau}}^{theo}-\mu_{\tau\bar{\tau}}^{exp}}{\delta_{\mu_{\tau\bar{\tau}}}})^2
+(\frac{\mu_{Z\gamma}^{theo}-\mu_{Z\gamma}^{exp}}{\delta_{\mu_{Z\gamma}}})^2
\end{eqnarray}
Firstly, we scan 14 free parameters in the following ranges: $\tan\beta\in(5,60)$, $\tan\beta'\in(1,1.5)$, $g_B\in(0.3,0.85)$, $g_{YB}\in(-0.45,-0.05)$,
$v_S\in(0.5,8)\mathrm{TeV}$, $\kappa\in(-2,3)$, $\lambda\in(-1.5,1.5)$, $\lambda_2\in(0.1,1.5)$,
$T_\kappa\in(-3,3)\mathrm{TeV}$, $T_\lambda\in(-2,2)\mathrm{TeV}$, $T_{\lambda_2}\in(-2,2)\mathrm{TeV}$, $M_{\tilde{Q}_3}\in(1.5,4)\mathrm{TeV}$, $M_{\tilde{t}}\in(1.5,4)\mathrm{TeV}$ and $M_2\in(0.1,1.5)\mathrm{TeV}$. As the SM-like Higgs boson mass satisfies $3\sigma$ experimental limits and the Higgs signal strengths $\mu_{\gamma\gamma, WW^*, ZZ^*, b\bar{b}, \tau\bar{\tau},Z\gamma}$ all satisfy $2\sigma$ experimental limits, we select the reasonable range for the free parameters after analyzing the scatter points in Fig.\ref{fig3}.

Fig.\ref{fig3}(a) presents the variation of the signal strength from Higgs boson to $Z$ boson and photon with the SM-like Higgs mass. The "rainbow" label corresponds to the value of $\chi^2$, with a maximum value around 18, which is below the $\chi^2$ critical value corresponding to the 14 degrees of freedom at $95\%$ CL(approximately 23.7). This implies that all scatter points in Fig.\ref{fig3} fall within $95\%$ CL. It is not difficult to find that the $h^0\rightarrow Z\gamma$ process in the NB-LSSM displays a deviation about $10\%$ from the SM, indicating the potentially significant loop-induced contributions of this model. The red star represents the best-fit point($\chi^2_{best}=5.176$) with Higgs boson mass around 125.26 GeV, which approximately aligns well with the observed SM-like Higgs boson mass. With the decrease of $\chi^2$, the mass of Higgs boson is closer to its center value, illustrating the strong experimental preference for a Higgs boson mass consistent with the SM-like Higgs boson.

The signal strengths of $h^0\rightarrow ZV$ processes with $V\in(\phi,J/\psi,\Upsilon(1S),\rho^0,\omega)$ can be found in Fig.\ref{fig3}(b) and (c). The values of signal strengths $\mu_{Z\phi}^{ggF}$ and $\mu_{ZJ/\psi}$ both increase with the increasing value of $\mu_{Z\gamma}^{ggF}$, revealing clear correlations among $\mu_{Z\gamma}^{ggF}$, $\mu_{Z\phi}^{ggF}$ and $\mu_{ZJ/\psi}^{ggF}$, as well as the similar correlations between $\mu_{Z\rho^0}^{ggF}$ and $\mu_{Z\omega}^{ggF}$. These correlations imply that these processes may be affected by one another and strongly constrained by the precision measurements of Higgs boson. Moreover, deviations exceeding $5\%$ from the SM expectations further support the presence of sizable NP contributions in the NB-LSSM.

As the mass of Higgs boson within $3\sigma$ experimental limits and the signal strengths of Higgs decays within $2\sigma$ experimental limits, we further display the scatter points of selected sensitive parameters in Fig.\ref{fig3}(d) and (e). Fig.\ref{fig3}(d) indicates that as the value of the parameter $\frac{\lambda*v_S}{\sqrt{2}}$ (which plays a role analogous to the MSSM $\mu$ parameter) continuously increases, the allowed range of the parameter $\kappa$ continuously decreases. That is, the scanning space of $|\kappa|$ decreases as $\frac{\lambda*v_S}{\sqrt{2}}$ increases. When $\kappa<0$, there are almost no scan points with $\lambda_2>0.5$, and as the value of $\frac{\lambda*v_S}{\sqrt{2}}$ increases, the scan points of $\lambda_2$ tend to approach -1.5. On the contrary, when $\kappa>0$, there are almost no scan points where $\lambda_2<-1.0$, and as the value of $\frac{\lambda*v_S}{\sqrt{2}}$ increases, the scan points of $\lambda_2$ tend more towards 1.5. Therefore, being compatible with the experimental measurements of Higgs signal strength, the couplings $\lambda$, $\lambda_2$, $\kappa$ and the VEV $v_S$ of Higgs singlet $S$ possess correlations with one another, and impose stringent constraints on the Yukawa structure in the Higgs sector.

In Fig.\ref{fig3}(e), the signal strength $\mu_{Z\gamma}^{ggF}$ tends to approach 1 when $\lambda$ increases, and its NP correction becomes weaker. Additionally, the scatter points with "rainbow" indicates that a smaller value of $v_S$ will make a greater NP contribution to the signal strength of rare Higgs decays. And as the value of parameter $\lambda$ increases, the values of $v_S $ consistent with the experimental constraints tend to decrease, leading to larger Higgs signal strengths. This indicates that there is a tightly correlated adjustment between these two parameters and exist an obvious influence on the rare Higgs decay observables.

To illustrate the parameter dependence more transparently, we next present one-dimensional scans for representative benchmark choices. We select some suitable parameter spaces for our numerical discussion.
\begin{eqnarray}
&&M_{2}=0.4 \mathrm{TeV},\; A_{t}=0.5\mathrm{TeV},\;A_{b}=0.2 \mathrm{TeV},\;A_{e}=0.3 \mathrm{TeV},\; M_{EE}=1.2 \mathrm{TeV}, \nonumber\\
&&M_{\tilde{Q}_3}=M_{\tilde{t}}=2.2 \mathrm{TeV},\;M_{\tilde{b}}=2 \mathrm{TeV},\; T_{\kappa}=M_{LL}=\Lambda=1\mathrm{TeV}, \; T_{\lambda}=0.9\mathrm{TeV},\nonumber\\
&&T_{\lambda_2}=-1.9\mathrm{TeV},\;v_S=1.9\mathrm{TeV},\; m_Z'=5.2\mathrm{TeV},\;\kappa=1.8,\; \lambda=0.4,\;\lambda_2=0.6,\nonumber\\
&&g_{YB}=-0.1,\; g_{B}=0.81,\; \tan \beta=45,\; \tan \beta^{\prime}=1.05.
\end{eqnarray}

\begin{figure}[t]
\centering
\includegraphics[width=0.4\textwidth]{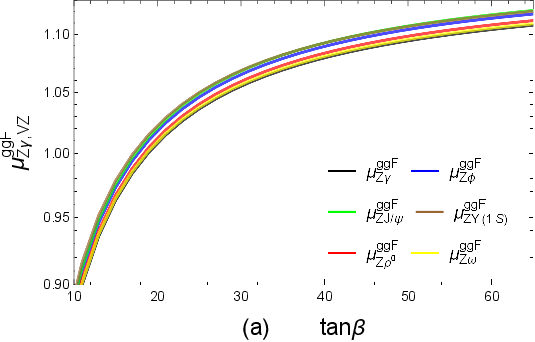}\\
\includegraphics[width=0.4\textwidth]{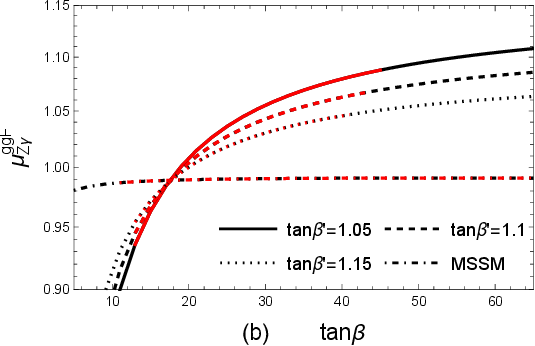}
\hspace{0.5cm}\includegraphics[width=0.4\textwidth]{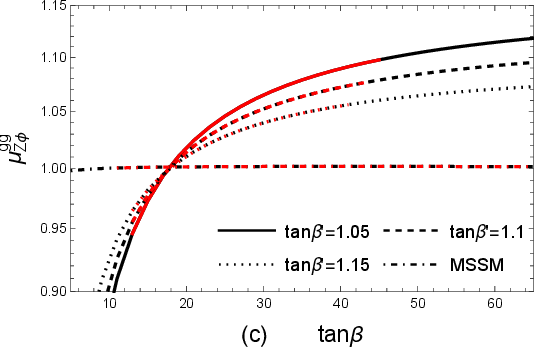}
\caption[]{The signal strengths of the rare Higgs decays change with parameters $\tan\beta$ and $\tan\beta'$, where the red lines meet the $3\sigma$ experimental limit of SM-like Higgs boson mass. The red lines in following figures satisfy the same constraints as here.}
\label{fig4}
\end{figure}
As shown in Fig.\ref{fig4}, we first analyzed the influence of parameters $\tan\beta$ and $\tan\beta'$ on the signal strengths. In Fig.\ref{fig4}(a), the signal strength $\mu_{Z\gamma,ZV}^{ggF}$ increases with the increased value of parameter $\tan\beta$, and the NB-LSSM framework can generate $10\%$ NP corrections. In addition, the curves of all these rare decays almost overlap, we will take $h^0\rightarrow Z\gamma,Z\phi$ processes as representative examples to carry out the discussion. In Fig.\ref{fig4}(b) and (c), the solid, dashed and dotted line correspond to $\tan\beta'=1.05,1.1,1.15$, respectively, while the dot dashed line represents the MSSM prediction. In the MSSM, the influence of $\tan\beta$ on the numerical results is mild, and the corresponding NP contribution is at the level of $2\%$. As the NP parameter in the NB-LSSM, $\tan\beta'$ is defined as $\tan\beta'=\frac{v_{\bar{\eta}}}{v_{\eta}}$, which can generate new corrections for rare decays by modifying all particle mass matrices and Higgs couplings. In particular, $\tan\beta'$ typically enlarges the allowed $\tan\beta$ and enhances the deviation of the signal strengths from 1, with maximal corrections reaching $10\%$. Therefore, a smaller $\tan\beta'$ generally yields larger NP contributions to rare Higgs decays in this model.

\begin{figure}[t]
\centering
\includegraphics[width=0.4\textwidth]{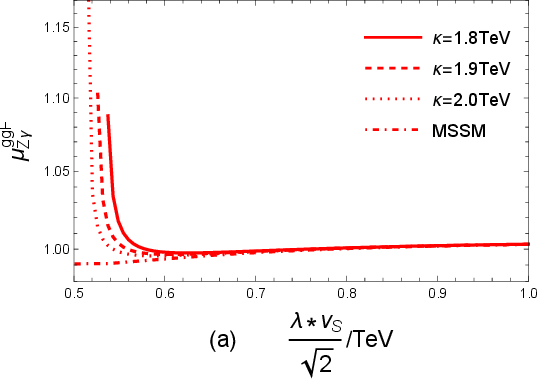}
\hspace{0.5cm}\includegraphics[width=0.4\textwidth]{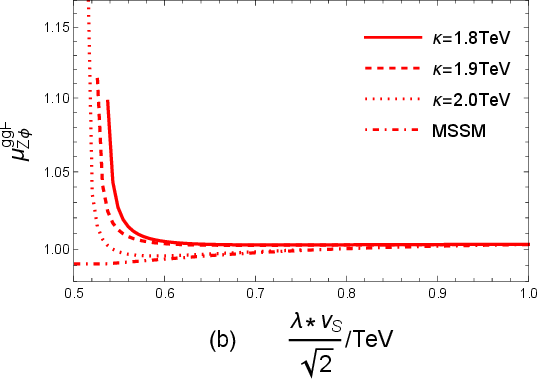}
\caption[]{The signal strengths of the rare Higgs decays change with parameters $\frac{\lambda*v_S}{\sqrt{2}}$ and $\kappa$.}
\label{fig5}
\end{figure}
Next, we study the effects from parameters $\frac{\lambda*v_S}{\sqrt{2}}$ and $\kappa$ to the numerical results in Fig.\ref{fig5}. The dot dashed curve represents the MSSM result, whose NP corrections are much smaller than those in the NB-LSSM. The solid, dashed and dotted line correspond to $\kappa=1.8,1.9, 2.0$, respectively. When the value of the parameter $\frac{\lambda*v_S}{\sqrt{2}}$ increases, the signal strength decreases rapidly and becomes relatively flat once $\frac{\lambda*v_S}{\sqrt{2}}$ being approximately greater than 0.6 TeV. When $\frac{\lambda*v_S}{\sqrt{2}}<0.6$ TeV and the signal strength takes a certain definite value, the larger $\kappa$ is, the smaller $\frac{\lambda*v_S}{\sqrt{2}}$ is. Moreover, increasing $\kappa$ enhances the NP contribution, with maximal deviations reaching $15\%$, highlighting the importance of the Yukawa coupling $\kappa$ in these rare decay channels.

\begin{figure}[t]
\centering
\includegraphics[width=0.4\textwidth]{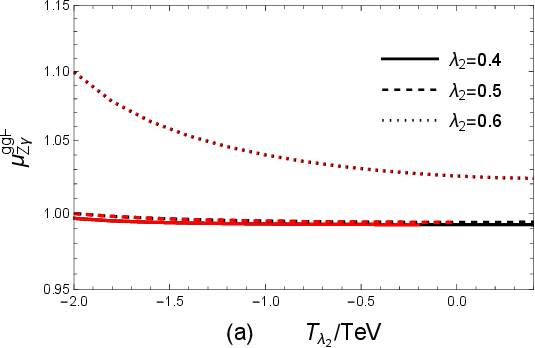}
\hspace{0.5cm}\includegraphics[width=0.4\textwidth]{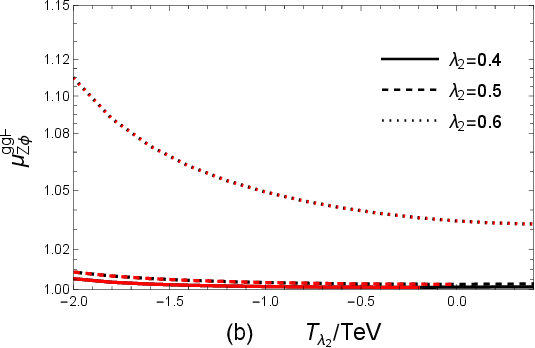}
\caption[]{The signal strengths of the rare Higgs decays change with parameters $T_{\lambda_2}$ and $\lambda_2$.}
\label{fig6}
\end{figure}
Then, the numerical results changing with trilinear soft parameter $T_{\lambda_2}$ are figured by Fig.\ref{fig6}. As the value of the parameter $|T_{\lambda_2}|$ decreases, the signal strengths of rare Higgs decays gradually weaken. When $\lambda_2$ is 0.4 or 0.5, the decline of the curve with the decrease of $|T_{\lambda_2}|$ is relatively small. However, when $\lambda_2$ is 0.6, the curve changes more pronounced with the decrease of $| T_{\lambda_2}|$, which indicates that larger $\lambda_2$ tends to induce larger NP corrections. In addition, the extent of the red curve also indicates that a broader parameter space of parameter $T_{\lambda_2}$ remains compatible with the current Higgs experimental constraints as $\lambda_2$ take a larger value. Overall, these features reflect a clear correlation between $T_{\lambda_2}$ and $\lambda_2$, which influence each other and jointly control the size of NP effects in the rare Higgs decays.

\begin{figure}[t]
\centering
\includegraphics[width=0.4\textwidth]{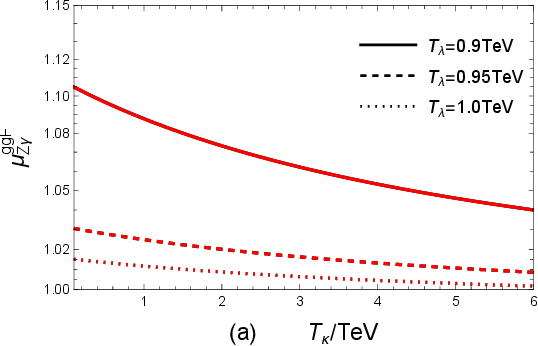}
\hspace{0.5cm}\includegraphics[width=0.4\textwidth]{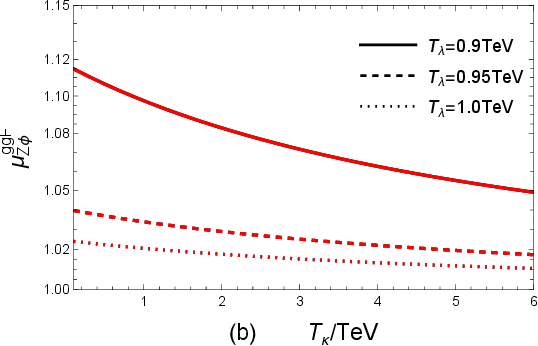}
\caption[]{The signal strengths of the rare Higgs decays change with parameters $T_{\kappa}$ and $T_{\lambda}$.}
\label{fig7}
\end{figure}
In Fig.\ref{fig7}, we analyze the influences from trilinear soft parameters $T_{\kappa}$ and $T_{\lambda}$, which affect the Higgs mass and couplings and thereby induce NP corrections. As the parameter $T_{\kappa}$ increases, the signal strength gradually decreases. A similar decreasing trend is observed as $T_{\lambda}$ increases. In addition, the curve for $T_{\lambda}=0.9\mathrm{TeV}$ is steeper than that for $T_{\lambda}=0.95\mathrm{TeV}$, while the curve corresponding to $T_{\lambda}=1.0\mathrm{TeV}$ is milder. Therefore, the smaller $T_{\lambda}$ is, the more obvious NP correction the $T_{\kappa}$ will have on the numerical results. If we take the smaller values of both $T_{\kappa}$ and $T_{\lambda}$, we can obtain a larger signal strength, highlighting the importance of these parameters for rare Higgs decay observables in the NB-LSSM.

\section{Discussion and conclusion}
In this paper, we study the rare decays $h^0\rightarrow Z\gamma,V Z$ with the meson $V\in(\phi,J/\psi,\Upsilon(1S),\rho^0,\omega)$ in the NB-LSSM. Compared with the MSSM, the NB-LSSM introduces three singlet Higgs superfields, which mix with the Higgs doublets, thereby modifying both the mass of the lightest Higgs boson and its couplings. These effects are controlled by the new parameters $\kappa$, $\lambda$, $\lambda_2$, $\tan\beta'$, $v_S$, $T_{\lambda}$, $T_{\kappa}$ and $T_{\lambda_2}$. As a consequence, the effective $h^0Z\gamma$ coupling in the NB-LSSM can receive sizable NP corrections through loop-induced contributions.

Requiring the SM-like Higgs mass lies within the $3\sigma$ experimental interval and that the Higgs signal strengths $\mu_{\gamma\gamma, WW^*, ZZ^*, b\bar{b}, \tau\bar{\tau},Z\gamma}^{ggF}$ all satisfy $2\sigma$ experimental bounds, we identify the viable parameter region after analyzing the scatter points. In particular, there is a certain correlation patterns among parameters $v_S$, $\lambda$, $\kappa$ and $\lambda_2$ shown by Fig.\ref{fig3}(d) and (e), which are tightly constrained by the Higgs measurements and strongly affect the numerical results. Additionally, we further perform one-dimensional scans to illustrate the dependence of signal strengths on parameters. We find that large $\tan\beta$, $\kappa$, $\lambda_2$, together with small $\tan\beta'$, $v_S$, $\lambda$, $T_{\lambda}$, $T_{\kappa}$, $T_{\lambda_2}$, can induce relatively large NP impacts on the results. When $\lambda_2(T_{\lambda})$ is large(small), $T_{\lambda_2}(T_{\kappa})$ becomes relatively sensitive and obviously influences the numerical results. Overall, the Higgs signal strengths of $h^0\rightarrow Z\gamma,V Z$ in the NB-LSSM can deviate by about $10\%$ from the SM. The results presented here provide a complementary probe for researching NP beyond SM through rare Higgs decays. With the continuous optimization of high-precision Higgs property measurements, rare decays such as $h^0\rightarrow Z\gamma,V Z$ will become increasingly sensitive probes of this scenario.

\begin{acknowledgments}
\indent
This work is supported by the Major Project of National Natural Science Foundation of China (NNSFC) (No. 12235008), the National Natural Science Foundation of China (NNSFC) (No. 12075074, No. 12075073), the Natural Science Foundation of Hebei province(No.A2022201022, No. A2023201041, No. A2022201017), the Natural Science Foundation of Hebei Education Department(No. QN2022173), the Project of the China Scholarship Council (CSC) (No. 202408130113). This work is also supported by Funda\c{c}\~{a}o para a Ci\^{e}ncia e a Tecnologia (FCT, Portugal) through the projects CFTP-FCT Unit UIDB/00777/2020, UIDP/00777/2020 and UID/00777/2025 (https://doi.org/10.54499/UID/00777/2025), which are partially funded through POCTI (FEDER), COMPETE, QREN and EU.
\end{acknowledgments}

\appendix
\section{The form factors}
The form factors are defined as
\begin{eqnarray}
&&A_{1 / 2}(\tau, \lambda) =I_{1}(\tau, \lambda)-I_{2}(\tau, \lambda),\;\;\; A_{0}(\tau, \lambda)  =I_{1}(\tau, \lambda),\nonumber\\
&&A_{1}(\tau, \lambda)  =c_{w}\left\{4\left(3-\frac{s_{w}^{2}}{c_{w}^{2}}\right) I_{2}(\tau, \lambda)+\left[\left(1+\frac{2}{\tau}\right) \frac{s_{w}^{2}}{c_{w}^{2}}-\left(5+\frac{2}{\tau}\right)\right] I_{1}(\tau, \lambda)\right\}, \nonumber\\
&&I_{1}(\tau, \lambda)  =\frac{\tau \lambda}{2(\tau-\lambda)}+\frac{\tau^{2} \lambda^{2}}{2(\tau-\lambda)^{2}}\left[g\left(\tau^{-1}\right)-g\left(\lambda^{-1}\right)\right]+\frac{\tau^{2} \lambda}{(\tau-\lambda)^{2}}\left[f\left(\tau^{-1}\right)-f\left(\lambda^{-1}\right)\right], \nonumber\\
&&I_{2}(\tau, \lambda)  =-\frac{\tau \lambda}{2(\tau-\lambda)}\left[g\left(\tau^{-1}\right)-g\left(\lambda^{-1}\right)\right], \nonumber\\
&&g(\tau)\hspace{-0.1cm}=\hspace{-0.15cm}\left\{\begin{array}{l}\hspace{-0.15cm}\arcsin^2\sqrt{\tau},\;\;\;\;\;\;\;\;\;\;\;\;\;\;\;\;\;\;\,\tau\le1\\
\hspace{-0.15cm}-{1\over4}\Big[\ln{1+\sqrt{1-1/\tau}\over1-\sqrt{1-1/\tau}}-i\pi\Big]^2,\tau>1\;\end{array}\right.,
f(\tau)\hspace{-0.1cm}=\hspace{-0.15cm}\left\{\begin{array}{l}
\hspace{-0.15cm}\sqrt{\tau^{-1}-1} \arcsin \sqrt{\tau}, \;\;\;\;\;\;\;\;\;\;\;\,\tau \geq 1 \\ \hspace{-0.15cm}\frac{\sqrt{1-\tau^{-1}}}{2}\left[\log \frac{1+\sqrt{1-1 / \tau}}{1-\sqrt{1-1 / \tau}}-i \pi\right], \tau<1 \end{array}\right..
\end{eqnarray}

\section{The coupling coefficients}
In the NB-LSSM, the concrete coupling coefficients contributed to the Higgs decays are specifically discussed as follows:

1. The Higgs-fermion-fermion contributions:
\begin{eqnarray}
&&{\cal L}^{\mathrm{NP}}_{{h^{k}uu}}=\frac{-im_u}{v}g_{h^{k}uu}=-\frac{i}{\sqrt{2}}\sum_{i=j_2=1}^{3}\sum_{j=j_1=1}^{3}  Y^*_{u,{j_1 j_2 }} U_{R,{j j_1}}^{u}  U_{L,{i j_2}}^{u}  Z_{{k 2}}^{H},
\nonumber\\
&&{\cal L}^{\mathrm{NP}}_{{h^{k}dd}}=\frac{-im_d}{v}g_{h^{k}dd}=-\frac{i}{\sqrt{2}}  \sum_{i=j_2=1}^{3}\sum_{j=j_1=1}^{3} Y^*_{d,{j_1 j_2}} U_{R,{j j_1}}^{d}  U_{L,{i j_2}}^{d}  Z_{{k 1}}^{H},
\nonumber\\
&&{\cal L}^{\mathrm{NP}}_{h^{k}l l}=\frac{-im_l}{v}g_{h^{k}ll}=-\frac{i}{\sqrt{2}}\sum_{i=j_2=1}^{3}\sum_{j=j_1=1}^{3} Y^*_{e,{j_1 j_2}} U_{R,{j j_1}}^{e}  U_{L,{i j_2}}^{e}  Z_{{k 1}}^{H},
\nonumber\\
&&{\cal L}^{\mathrm{NP}}_{{h^{k}\chi_i^+\chi_j^-}}=\frac{-ie}{2s_w}(g_{h^{k}\chi_i^+\chi_j^-}+\tilde{g}_{h^{k}\chi_i^+\chi_j^-}\gamma^5)
\nonumber\\
&&\hspace{1.5cm}=-\frac{i}{\sqrt{2}} \sum_{i,j=1}^2 \Big[\Big(g_2U^*_{{j 1}} V^*_{{i 2}} Z_{{k 2}}^{H}  + U^*_{{j 2}}(g_2 V^*_{{i 1}} Z_{{k 1}}^{H}+\lambda V^*_{{i 2}} Z_{{k 5}}^{H}) \Big)P_L\nonumber\\
&&\hspace{1.5cm}+\Big(g_2U_{{i 1}} V_{{j 2}} Z_{{k 2}}^{H}  + U_{{i 2}}(g_2 V_{{j 1}} Z_{{k 1}}^{H}+\lambda^* V_{{j 2}} Z_{{k 5}}^{H}) \Big)P_R\Big].
\end{eqnarray}
2. The Higgs-$W(Z)$-$W(Z)$ boson contributions:
\begin{eqnarray}
&&{\cal L}^{\mathrm{NP}}_{h^{k}WW}=ig_2m_Wg_{h^{k}WW}=\frac{i}{2}g_2^2\Big(v_d Z_{{k 1}}^{H}  + v_u Z_{{k 2}}^{H} \Big),
\nonumber\\&&{\cal L}^{\mathrm{NP}}_{h^kZZ}=i\frac{2m_Z^2}{v}g_{h^kZZ}=i\Big[\frac{1}{2} \Big(v_d (g_1 c'_w  s_w   + g_2 c_w  c'_w  \hspace{-0.1cm} - \hspace{-0.1cm}g_{Y B} s'_w  )^{2} Z_{{k 1}}^{H}+v_u \Big(g_1 c'_w  s_w  \hspace{-0.1cm} +\hspace{-0.1cm} g_2 c_w  c'_w  \nonumber\\
&&\hspace{1.2cm}-\hspace{-0.1cm} g_{Y B} s'_w  )^{2} Z_{{k 2}}^{H} +4 (- g_{B} s'_w )^{2} (v_{\bar{\eta}} Z_{{k 4}}^{H}  + v_{\eta} Z_{{k 3}}^{H} )\Big)\Big].
\end{eqnarray}

3. The Higgs-scalar-scalar contributions:
\begin{eqnarray}
&&{\cal L}^{\mathrm{NP}}_{h^k\tilde{U}\tilde{U}}=\frac{-im_W^2}{v}g_{h^{k}\tilde{U}\tilde{U}}
\nonumber \\
&&=\frac{i}{12}\sum_{i,j=1}^6\Big[ 6 \Big( v_S\lambda^*\sum_{j_2=1}^{3}Z^{U,*}_{j j_2} \sum_{j_1=1}^{3}Y_{u,{j_1 j_2}} Z_{{i 3 + j_1}}^{U}   Z_{{k 1}}^{H} +\hspace{-0.1cm}v_S\lambda\hspace{-0.1cm} \sum_{j_2=1}^{3}\hspace{-0.1cm}\sum_{j_1=1}^{3}\hspace{-0.1cm}Y^*_{u,{j_1 j_2}} Z^{U,*}_{j 3 + j_1}  Z_{{i j_2}}^{U}  Z_{{k 1}}^{H}
\nonumber \\
&&+v_u\lambda^*\sum_{j_2=1}^{3}Z^{U,*}_{j j_2} \sum_{j_1=1}^{3}Y_{u,{j_1 j_2}} Z_{{i 3 + j_1}}^{U}   Z_{{k 5}}^{H}+\hspace{-0.1cm}v_u\lambda\hspace{-0.1cm} \sum_{j_2=1}^{3}\hspace{-0.1cm}\sum_{j_1=1}^{3}\hspace{-0.1cm}Y^*_{u,{j_1 j_2}} Z^{U,*}_{j 3 + j_1}  Z_{{i j_2}}^{U}  Z_{{k 5}}^{H}\hspace{-0.1cm}\nonumber \\
&&-\hspace{-0.1cm} \Big(\hspace{-0.1cm}\sqrt{2} \hspace{-0.1cm}\sum_{j_2=1}^{3}\hspace{-0.1cm}Z^{U,*}_{j j_2} \sum_{j_1=1}^{3}\hspace{-0.1cm}Z_{{i 3 + j_1}}^{U} T_{u,{j_1 j_2}}  \hspace{-0.1cm} +\hspace{-0.1cm}\sqrt{2} \hspace{-0.1cm}\sum_{j_2=1}^{3}\sum_{j_1=1}^{3}\hspace{-0.1cm}Z^{U,*}_{j 3 + j_1} T^*_{u,{j_1 j_2}}  Z_{{i j_2}}^{U}  \nonumber \\
&&+2 v_u (\sum_{j_3=1}^{3}Z^{U,*}_{j 3 + j_3} \sum_{j_2=1}^{3}\sum_{j_1=1}^{3}Y^*_{u,{j_3 j_1}} Y_{u,{j_2 j_1}}  Z_{{i 3 + j_2}}^{U}   + \sum_{j-3=1}^{3}\sum_{j_2=1}^{3}Z^{U,*}_{j j_2} \sum_{j_1=1}^{3}Y^*_{u,{j_1 j_3}} Y_{u,{j_1 j_2}}   Z_{{i j_3}}^{U} )\Big)Z_{{k 2}}^{H} \Big)\nonumber \\
&&+\hspace{-0.1cm}\sum_{j_1=1}^{3}\hspace{-0.1cm}Z^{U,*}_{j 3 + j_1} Z_{{i 3 + j_1}}^{U}  \Big(\hspace{-0.1cm}-\hspace{-0.1cm} \Big(4 g_{1}^{2} \hspace{-0.1cm}+\hspace{-0.1cm} g_{Y B}(4 g_{Y B} \hspace{-0.1cm} +\hspace{-0.1cm} g_{B})\Big)v_d Z_{{k 1}}^{H} \hspace{-0.1cm}+\hspace{-0.1cm}\Big(4 g_{1}^{2} \hspace{-0.1cm}+\hspace{-0.1cm} g_{Y B}(4 g_{Y B} \hspace{-0.1cm} +\hspace{-0.1cm} g_{B})\Big)v_u Z_{{k 2}}^{H} \nonumber \\
&&-2( 4 g_{Y B} g_{B} \hspace{-0.1cm} +\hspace{-0.1cm} g_{B}^{2})(\hspace{-0.1cm}-\hspace{-0.1cm} v_{\bar{\eta}} Z_{{k 4}}^{H} \hspace{-0.1cm} +\hspace{-0.1cm} v_{\eta} Z_{{k 3}}^{H} )\Big)\hspace{-0.1cm}+\hspace{-0.1cm}\sum_{j_1=1}^{3}\hspace{-0.1cm}Z^{U,*}_{j j_1} Z_{{i j_1}}^{U}  \Big((\hspace{-0.1cm}-\hspace{-0.1cm}3 g_{2}^{2} \hspace{-0.1cm}+\hspace{-0.1cm} g_{Y B} g_{B} \hspace{-0.1cm} +\hspace{-0.1cm} g_{1}^{2}\hspace{-0.1cm} +\hspace{-0.1cm} g_{Y B}^{2})v_d Z_{{k 1}}^{H} \nonumber \\
&&-(-3 g_{2}^{2} + g_{Y B} g_{B}  + g_{1}^{2} + g_{Y B}^{2})v_u Z_{{k 2}}^{H} +2 (g_{Y B} g_{B}  + g_{B}^{2})(- v_{\bar{\eta}} Z_{{k 4}}^{H}  + v_{\eta} Z_{{k 3}}^{H})\Big)\Big],\nonumber \\
&&
{\cal L}^{\mathrm{NP}}_{h^k\tilde{D}\tilde{D}}=\frac{-im_W^2}{v}g_{h^{k}\tilde{D}\tilde{D}}
\nonumber \\
&&=\hspace{-0.1cm}\frac{i}{12} \sum_{i,j=1}^6\hspace{-0.1cm}\Big[ \hspace{-0.1cm}-\hspace{-0.1cm}6 (\sqrt{2} \sum_{j_2=1}^{3}\hspace{-0.1cm}Z^{D,*}_{j j_2} \sum_{j_1=1}^{3}\hspace{-0.1cm}Z_{{i 3 + j_1}}^{D} T_{d,{j_1 j_2}}   Z_{{k 1}}^{H} \hspace{-0.1cm}+\hspace{-0.1cm}\sqrt{2} \sum_{j_2=1}^{3}\hspace{-0.1cm}\sum_{j_1=1}^{3}\hspace{-0.1cm}Z^{D,*}_{j 3 + j_1} T^*_{d,{j_1 j_2}}  Z_{{i j_2}}^{D}  Z_{{k 1}}^{H} \nonumber \\
&&+2 v_d \sum_{j_3=1}^{3}Z^{D,*}_{j 3 + j_3} \sum_{j_2=1}^{3}\sum_{j_1=1}^{3}Y^*_{d,{j_3 j_1}} Y_{d,{j_2 j_1}}  Z_{{i 3 + j_2}}^{D}   Z_{{k 1}}^{H} +2 v_d \sum_{j_3=1}^{3}\sum_{j_2=1}^{3}Z^{D,*}_{j j_2} \sum_{j_1=1}^{3}Y^*_{d,{j_1 j_3}} Y_{d,{j_1 j_2}}   Z_{{i j_3}}^{D}  Z_{{i 1}}^{H} \nonumber \\
&&- v_S\lambda^* \sum_{j_2=1}^{3}Z^{D,*}_{j j_2} \sum_{j_1=1}^{3}Y_{d,{j_1 j_2}} Z_{{i 3 + j_1}}^{D}   Z_{{k 2}}^{H} - v_S\lambda \sum_{j_2=1}^{3}\sum_{j_1=1}^{3}Y^*_{d,{j_1 j_2}} Z^{D,*}_{j 3 + j_1}  Z_{{i j_2}}^{D}  Z_{{k 2}}^{H} \nonumber \\
&&- v_u\lambda^* \sum_{j_2=1}^{3}Z^{D,*}_{j j_2} \sum_{j_1=1}^{3}Y_{d,{j_1 j_2}} Z_{{i 3 + j_1}}^{D}   Z_{{k 5}}^{H} - v_u\lambda \sum_{j_2=1}^{3}\sum_{j_1=1}^{3}Y^*_{d,{j_1 j_2}} Z^{D,*}_{j 3 + j_1}  Z_{{i j_2}}^{D}  Z_{{k 5}}^{H} )\nonumber \\
&&+\hspace{-0.1cm}\sum_{j_1=1}^{3}\hspace{-0.1cm}Z^{D,*}_{j 3 + j_1} Z_{{i 3 + j_1}}^{D}  \Big(\Big(2 g_{1}^{2} \hspace{-0.1cm}+\hspace{-0.1cm} g_{Y B}(2 g_{Y B}  \hspace{-0.1cm}-\hspace{-0.1cm} g_{B})\Big)v_d Z_{{k 1}}^{H} \hspace{-0.1cm}+\hspace{-0.1cm}\Big(\hspace{-0.1cm}-\hspace{-0.1cm}2 g_{1}^{2} \hspace{-0.1cm}+\hspace{-0.1cm} g_{Y B} (\hspace{-0.1cm}-\hspace{-0.1cm}2 g_{Y B} \hspace{-0.1cm} + \hspace{-0.1cm} g_{B})\Big)v_u Z_{{k 2}}^{H} \nonumber\\
&&+2(2 g_{Y B} g_{B}  \hspace{-0.1cm}-\hspace{-0.1cm} g_{B}^{2})(\hspace{-0.1cm}- \hspace{-0.1cm} v_{\bar{\eta}} Z_{{k 4}}^{H} \hspace{-0.1cm} +\hspace{-0.1cm} v_{\eta} Z_{{k 3}}^{H} )\Big)\hspace{-0.1cm}+\hspace{-0.1cm}\sum_{j_1=1}^{3}\hspace{-0.1cm}Z^{D,*}_{j j_1} Z_{{i j_1}}^{D}  \Big((3 g_{2}^{2} \hspace{-0.1cm}+\hspace{-0.1cm} g_{Y B} g_{B} \hspace{-0.1cm} +\hspace{-0.1cm} g_{1}^{2}\hspace{-0.1cm} +\hspace{-0.1cm} g_{Y B}^{2})v_d Z_{{k 1}}^{H} \nonumber \\
&&-(3 g_{2}^{2} + g_{Y B} g_{B}  + g_{1}^{2} + g_{Y B}^{2})v_u Z_{{k 2}}^{H} +2( g_{Y B} g_{B}  + g_{B}^{2})(- v_{\bar{\eta}} Z_{{k 4}}^{H}  + v_{\eta} Z_{{k 3}}^{H})\Big)\Big],\nonumber \\
&&{\cal L}^{\mathrm{NP}}_{h^k\tilde{L}\tilde{L}}=\frac{-im_W^2}{v}g_{h^{k}\tilde{L}\tilde{L}}
\nonumber \\
&&=\frac{i}{4}\sum_{i,j=1}^6\Big[ \hspace{-0.1cm}- \hspace{-0.1cm}2 (\sqrt{2}  \hspace{-0.1cm}\sum_{j_2=1}^{3}\hspace{-0.1cm}Z^{E,*}_{j j_2}\hspace{-0.1cm} \sum_{j_1=1}^{3}\hspace{-0.1cm}Z_{{i 3 + j_1}}^{E} T_{e,{j_1 j_2}}   Z_{{k 1}}^{H}  \hspace{-0.1cm}+ \hspace{-0.1cm}\sqrt{2} \sum_{j_2=1}^{3}\sum_{j_1=1}^{3}Z^{E,*}_{j 3 + j_1} T^*_{e,{j_1 j_2}}  Z_{{i j_2}}^{E}  Z_{{k 1}}^{H} \nonumber \\
&&+2 v_d \sum_{j_3=1}^{3}Z^{E,*}_{j 3 + j_3} \sum_{j_2=1}^{3}\sum_{j_1=1}^{3}Y^*_{e,{j_3 j_1}} Y_{e,{j_2 j_1}}  Z_{{i 3 + j_2}}^{E}   Z_{{k 1}}^{H} +2 v_d \sum_{j_3=1}^{3}\sum_{j_2=1}^{3}Z^{E,*}_{j j_2} \sum_{j_1=1}^{3}Y^*_{e,{j_1 j_3}} Y_{e,{j_1 j_2}}   Z_{{i j_3}}^{E}  Z_{{k 1}}^{H} \nonumber \\
&&- v_S\lambda^* \sum_{j_2=1}^{3}Z^{E,*}_{j j_2} \sum_{j_1=1}^{3}Y_{e,{j_1 j_2}} Z_{{i 3 + j_1}}^{E}   Z_{{k 2}}^{H} - v_S\lambda \sum_{j_2=1}^{3}\sum_{j_1=1}^{3}Y^*_{e,{j_1 j_2}} Z^{E,*}_{j 3 + j_1}  Z_{{i j_2}}^{E}  Z_{{k 2}}^{H}\nonumber \\
&&- v_u\lambda^* \sum_{j_2=1}^{3}Z^{E,*}_{j j_2} \sum_{j_1=1}^{3}Y_{e,{j_1 j_2}} Z_{{i 3 + j_1}}^{E}   Z_{{k 5}}^{H} - v_u\lambda \sum_{j_2=1}^{3}\sum_{j_1=1}^{3}Y^*_{e,{j_1 j_2}} Z^{E,*}_{j 3 + j_1}  Z_{{i j_2}}^{E}  Z_{{k 5}}^{H})\nonumber \\
&&+\sum_{j_1=1}^{3}\hspace{-0.1cm}Z^{E,*}_{j 3 + j_1} Z_{{i 3 + j_1}}^{E}  \Big(\Big(2 g_{1}^{2} \hspace{-0.1cm}+\hspace{-0.1cm} g_{Y B}(2 g_{Y B} \hspace{-0.1cm} +\hspace{-0.1cm} g_{B})\Big)v_d Z_{{k 1}}^{H} \hspace{-0.1cm}- \hspace{-0.1cm}\Big(2 g_{1}^{2} \hspace{-0.1cm}+\hspace{-0.1cm} g_{Y B}(2 g_{Y B} \hspace{-0.1cm} +\hspace{-0.1cm} g_{B})\Big)v_u Z_{{k 2}}^{H} \nonumber \\
&&+2 (2 g_{Y B} g_{B}  \hspace{-0.1cm}+ \hspace{-0.1cm}g_{B}^{2})(\hspace{-0.1cm}-\hspace{-0.1cm} v_{\bar{\eta}} Z_{{k 4}}^{H} \hspace{-0.1cm} +\hspace{-0.1cm} v_{\eta} Z_{{k 3}}^{H} )\hspace{-0.1cm}\Big)\hspace{-0.1cm}+\hspace{-0.1cm}\sum_{j_1=1}^{3}\hspace{-0.1cm}Z^{E,*}_{j j_1} Z_{{i j_1}}^{E}  \Big(\hspace{-0.1cm}-\hspace{-0.1cm} (\hspace{-0.1cm}-\hspace{-0.1cm} g_{2}^{2} \hspace{-0.1cm} +\hspace{-0.1cm} g_{Y B} g_{B} \hspace{-0.1cm} +\hspace{-0.1cm} g_{1}^{2} \hspace{-0.1cm}+\hspace{-0.1cm} g_{Y B}^{2})v_d Z_{{k 1}}^{H}  \nonumber \\
&&+(- g_{2}^{2}  + g_{Y B} g_{B}  + g_{1}^{2} + g_{Y B}^{2})v_u Z_{{k 2}}^{H}-2 ( g_{Y B} g_{B}  + g_{B}^{2})(- v_{\bar{\eta}} Z_{{k 4}}^{H}  + v_{\eta} Z_{{k 3}}^{H})\Big)\Big],\nonumber \\
&&{\cal L}^{\mathrm{NP}}_{h^kH^{\pm}H^{\pm}}=\frac{-im_W^2}{vc_w^2}g_{h^{k}H^{\pm}H^{\pm}}
\nonumber \\
&&=\frac{i}{4} \sum_{i,j=1}^2\Big[ -2 g_{YB} g_{B}(- v_{\bar{\eta}} Z_{{k 4}}^{H} + v_{\eta} Z_{{k 3}}^{H})(Z_{{j 1}}^{+} Z_{{i 1}}^{+} - Z_{{j 2}}^{+} Z_{{i 2}}^{+})\nonumber\\
&&+\lambda_2\lambda^*(v_{\eta} Z_{{k 4}}^{H} + v_{\bar{\eta}} Z_{{k 3}}^{H})(Z_{{j 2}}^{+} Z_{{i 1}}^{+} +Z_{{j 1}}^{+} Z_{{i 2}}^{+})+Z_{{k 5}}^{H}\big(2v_S\lambda^*\lambda(Z_{{j 1}}^{+} Z_{{i 1}}^{+} \nonumber\\
&&+ Z_{{j 2}}^{+} Z_{{i 2}}^{+})+(2v_S\lambda^*\kappa+\sqrt{2}T_\lambda^*)Z_{{j 2}}^{+} Z_{{i 1}}^{+} +(2v_S\lambda\kappa^*+\sqrt{2}T_\lambda)Z_{{j 1}}^{+} Z_{{i 2}}^{+}\big)\nonumber\\
&&- Z_{{k 1}}^{H} \Big(Z_{{j 2}}^{+} \Big(-(-g_{2}^{2}+g_{1}^{2}+g_{Y B}^{2})v_d Z_{{i 2}}^{+}+(-2|\lambda|^2+g_{2}^{2}) v_u Z_{{i 1}}^{+} \Big)\nonumber \\
&&\hspace{0.8cm}+Z_{{j 1}}^{+} \Big((g_{1}^{2}+g_{Y B}^{2}+g_{2}^{2})v_d Z_{{i 1}}^{+}+(-2|\lambda|^2+g_{2}^{2}) v_u Z_{{i 2}}^{+}\Big)\Big)\nonumber\\
&&+Z_{{k 2}}^{H}\Big(Z_{{j 1}}^{+} \Big((-g_{2}^{2}+ g_{1}^{2}+g_{Y B}^{2})v_u Z_{{i 1}}^{+}-(-2|\lambda|^2+g_{2}^{2}) v_d Z_{{i 2}}^{+} \Big)\nonumber \\
&&\hspace{0.8cm}-Z_{{j 2}}^{+}\Big((g_{1}^{2}+g_{Y B}^{2}+g_{2}^{2}))v_u Z_{{i 2}}^{+}+ (-2|\lambda|^2+g_{2}^{2}) v_d Z_{{i 1}}^{+} \Big)\Big].
\end{eqnarray}

4. The $Z$ boson-fermion-fermion contributions:
\begin{eqnarray}
&&{\cal L}^{\mathrm{NP}}_{Zdd}=\frac{-ie}{s_wc_w}(g_{Zdd}\gamma^\mu+\tilde{g}_{Zdd}\gamma^\mu\gamma^5)
\nonumber\\
&&=\frac{i}{6}\gamma^\mu[(3g_2 c_w c'_w+g_1 c'_w s_w-(g_B+g_{YB})s'_w) P_L-(2g_1 c'_w s_w+(-2g_{YB}+g_B)s'_w)P_R],
\nonumber\\
&&{\cal L}^{\mathrm{NP}}_{{Zll}}=\frac{-ie}{s_wc_w}(g_{Zll}\gamma^\mu+\tilde{g}_{Zll}\gamma^\mu\gamma^5)\nonumber\\
&&=\frac{i}{2}\gamma^\mu[(-g_1 c'_w s_w+g_2 c_w c'_w+(g_B+g_{YB})s'_w) P_L-(2g_1 c'_w s_w-(2g_{YB}+g_B)s'_w) P_R],
\nonumber\\
&&{\cal L}^{\mathrm{NP}}_{{Zuu}}=\frac{-ie}{s_wc_w}(g_{Zuu}\gamma^\mu+\tilde{g}_{Zuu}\gamma^\mu\gamma^5)\nonumber\\
&&=-\frac{i}{6}\gamma^\mu[(3g_2 c_w c'_w-g_1 c'_w s_w+(g_B+g_{YB})s'_w) P_L-(4g_1 c'_w s_w-(4g_{YB}+g_B)s'_w)P_R],
\nonumber\\
&&{\cal L}^{\mathrm{NP}}_{{Z\chi_i^+\chi_j^-}}=\frac{-ie}{2s_wc_w}(g_{Z\chi_i^+\chi_j^-}\gamma^\mu+\tilde{g}_{Z\chi_i^+\chi_j^-}\gamma^\mu\gamma^5)
\nonumber\\
&&\hspace{1.5cm}=\frac{i}{2} \sum_{i,j=1}^2\Big[\Big(2 g_2 c_w  c'_w U^*_{j 1} U_{{i 1}}
+ (- g_1 c'_w  s'_w   + g_2 c_w  c'_w   + g_{Y B} s'_w  )U^*_{j 2}U_{{i 2}} \Big)\gamma_{\mu}P_L\nonumber\\
&&\hspace{1.5cm}+  \Big(2 g_2 c_w  c'_w V^*_{i 1} V_{{j 1}} +(- g_1 c'_w  s_w   + g_2 c_w  c'_w   + g_{Y B} s'_w  )V^*_{i 2} V_{{j 2}} \Big)\gamma_{\mu}P_R\Big].
\end{eqnarray}

5. The $Z$-$W$-$W$ boson contributions:
\begin{eqnarray}
{\cal L}^{\mathrm{NP}}_{ZWW}=-ig_2c_wg_{ZWW}=-ig_2c_wc'_w.
\end{eqnarray}

6. The $Z$ boson-scalar-scalar contributions:
\begin{eqnarray}
&&{\cal L}^{\mathrm{NP}}_{Z\tilde{U}\tilde{U}}=\frac{-ie}{s_wc_w}g_{Z\tilde{U}\tilde{U}}
=-\frac{i}{6} \sum_{i,j=1}^6 \delta_{\alpha \beta} \Big[\Big(3 g_2 c_w  c'_w   -g_1 c'_w  s_w   + \Big(g_{Y B} + g_{B}\Big)s'_w  \Big)\sum_{a=1}^{3}Z^{U,*}_{i a} Z_{{j a}}^{U}  \nonumber \\
&&\hspace{1.5cm}+\Big(- 4 g_1 c'_w  s_w   + \Big(4 g_{Y B}  + g_{B}\Big)s'_w  \Big)\sum_{a=1}^{3}Z^{U,*}_{i 3 + a} Z_{{j 3 + a}}^{U}  \Big],
\nonumber \\
&&{\cal L}^{\mathrm{NP}}_{Z\tilde{D}\tilde{D}}=\frac{-ie}{s_wc_w}g_{Z\tilde{D}\tilde{D}}
=\frac{i}{6} \sum_{i,j=1}^6 \delta_{\alpha \beta} \Big[\Big(3 g_2 c_w  c'_w   + g_1c'_w  s_w   - \Big(g_{Y B} + g_{B}\Big)s'_w  \Big)\sum_{a=1}^{3}Z^{D,*}_{i a} Z_{{j a}}^{D}  \nonumber \\
&&\hspace{1.5cm}+\Big(-2 g_1  c'_w  s_w   + \Big(2 g_{Y B}  - g_{B} \Big)s'_w  \Big)\sum_{a=1}^{3}Z^{D,*}_{i 3 + a} Z_{{j 3 + a}}^{D}  \Big],
\nonumber \\
&&{\cal L}^{\mathrm{NP}}_{Z\tilde{L}\tilde{L}}=\frac{-ie}{s_wc_w}g_{Z\tilde{L}\tilde{L}}
=\frac{i}{2}\sum_{i,j=1}^6 \Big[\Big(- g_1 c'_w  s_w   + g_2 c_w  c'_w   + \Big(g_{Y B} + g_{B}\Big)s'_w  \Big)\sum_{a=1}^{3}Z^{E,*}_{i a} Z_{{j a}}^{E}  \nonumber \\
 &&\hspace{1.5cm}+\Big(- 2 g_1  c'_w  s_w   + \Big(2 g_{Y B}  + g_{B}\Big)s'_w  \Big)\sum_{a=1}^{3}Z^{E,*}_{i 3 + a} Z_{{j 3 + a}}^{E}  \Big],
\nonumber\\
&&{\cal L}^{\mathrm{NP}}_{ZH^{\pm}H^{\pm}}=\frac{ie}{2s_wc_w}g_{ZH^{\pm}H^{\pm}}
=\frac{i}{2} \delta_{i j} (- g_1 c'_w  s_w   + g_2 c_w  c'_w   + g_{Y B} s'_w  ).
\end{eqnarray}

\end{document}